\documentclass[prd,nofootinbib,floats, superscriptaddress,eqsecnum, tightenlines, showpacs,letterpaper,12pt]{revtex4}

\usepackage{amsfonts,amssymb,amstext,amsmath, amsthm,bm,slashed}
\usepackage[dvips]{graphicx}
\usepackage{mathtools}
\usepackage{xcolor}
\usepackage[hidelinks]{hyperref}
\usepackage{tabularx}
\theoremstyle{definition}

\theoremstyle{plain}

\usepackage{cleveref}
\usepackage{relsize}
\usepackage[utf8]{inputenc}

\newcommand{\be}{\begin{equation}}
\newcommand{\ee}{\end{equation}}
\newcommand{\barray}{\begin{array}}
\newcommand{\earray}{\end{array}}
\newcommand{\bea}{\begin{eqnarray}}
\newcommand{\eea}{\end{eqnarray}}
\newcommand{\bs}{\begin{subequations}}
\newcommand{\es}{\end{subequations}}
\newcommand{\beal}{\begin{align}}
\newcommand{\eeal}{\end{align}}

\newcommand{\Ref}[1]{(\ref{#1})}

\def\bs{\bar{\sigma}}

\def\pa{\partial}

\newcommand{\rd}{\mathrm{d}}
\newcommand{\p}{\partial}

\newcommand*\overbar[1]{%
\hbox{%
\vbox{%
  \hrule height 0.5pt 
  \kern0.4ex
  \hbox{%
    \kern 0em
    \ensuremath{#1}%
    \kern 0em
  }%
}%
}%
} 

\def \Tr {\mathrm{Tr}}
\def \tr{\mathrm{tr}}

\usepackage{bbm}

\allowdisplaybreaks

\newcommand{\nnn}{\nonumber\\}

\def\l{{{\ell}}}

\def\Lb{{\bar L}}
\def\const{\text{const.}}
\def\rd{{\mathrm{d}}}

\def\bb{{\bar\beta}}
\renewcommand{\L}{{\mathcal{L}}}

\newcommand{\LL}{{\mathfrak{L}}}

\newcommand{\eb}{{\epsilon_{\mathrm{B}}}}
\renewcommand{\es}{{\epsilon_{\mathrm{S}}}}

\begin{document}

\title{Null Conservation Laws for Gravity}

\author{Florian Hopfmüller}
\email{fhopfmueller@perimeterinstitute.ca}
\affiliation{Perimeter Institute for Theoretical Physics\\
31 Caroline St. N, N2L 2Y5, Waterloo ON, Canada}

\author{Laurent Freidel}
\email{lfreidel@perimeterinstitute.ca}
\affiliation{Perimeter Institute for Theoretical Physics\\
31 Caroline St. N, N2L 2Y5, Waterloo ON, Canada}

\date{\today}

\begin{abstract}
We give a full analysis of the conservation along null surfaces of generalized energy and super-momenta, for gravitational systems enclosed by a finite boundary.
In particular we interpret the conservation equations in a canonical manner, revealing a notion of symplectic potential and a boundary current intrinsic to null surfaces. This generalizes similar analyses done at asymptotic infinity or on horizons.
\end{abstract}

\maketitle

\newpage
\thispagestyle{empty}
\newpage

\section{Introduction}

Gauge theories in regions with boundaries at finite or infinite distance have attracted much attention recently. 
Some of the key recent developments \cite{Strominger:2017zoo} which relate soft theorems with asymptotic symmetries have to do with the canonical structure and conservation laws arising at null infinity \cite{Ashtekar:1981bq}. A central feature here is the presence of soft modes governing the asymptotic dynamics of massive or massless particles. Soft modes allow to unify seemingly different concepts such as  infrared dressings, memory effects and asymptotic symmetries.

A related development comes from the more general study 
of gravity and gauge theories in the presence of finite boundaries.
It has been established \cite{Balachandran:1994up,Donnelly:2016auv} that gauge theories in the presence of finite boundaries admit additional edge mode degrees of freedom, which are necessary for constructing the global Hilbert space associated with the gluing of  subregions. These edge modes transform under a new set of boundary symmetries, which form an infinite dimensional  group \cite{Donnelly:2016auv, Speranza:2017gxd, Balachandran:2013wsa}.

We conjecture that when the finite boundary is moved to infinity the boundary symmetries become asymptotic symmetries and include the BMS symmetries in asymptotically flat spacetimes, while edge modes are identified with soft modes. Edge modes thus provide the hope of finding unifying descriptions of phenomena in the fields of entanglement entropy, soft theorems, infrared dressings, memory effects, black hole thermodynamics, and holography.

One of the key features of the asymptotic symmetry story is the presence of an asymptotic dynamics, generalizing Bondi mass evolution \cite{Bondi:1990zza}, that governs the evolution of the asymptotic symmetry generators. The asymptotic dynamics play a key role in the gravitational memory effect \cite{Christodoulou:1991cr}. In this work we are interested in generalizing these results to the case of finite boundaries.

We give an analysis of the evolution of symmetry generators along a general null hypersurface in a general Einstein spacetime. We clarify the canonical nature of these conservations laws, relating them with the recent construction of the null symplectic potential \cite{Epp:1995uc, Parattu:2016trq, Hopfmuller:2016scf, DePaoli:2017sar, Wieland:2017zkf}.
In the case of null infinity, it has proven fruitful to take a canonical viewpoint, and view the gravitational constraints as conservation equations for quantities which have a canonical meaning (e.g., \cite{He:2014laa}). A similar analysis was missing for finite boundaries and we provide a complete description of the canonical evolution along null surfaces together with a detailed understanding of the canonical role of the constraints on a null surface. This analysis is key to eventually understanding the dynamics of null gravity edge modes at finite distance
\cite{Donnelly:2016auv, Speranza:2017gxd}.

Our study can be viewed as an extension of the works \cite{Price:1986yy, Torre:1985rw, 0264-9381-10-4-012, hayward1993dual, Reisenberger:2012zq} which have studied  gravitational constraints on a null surface. In addition, our analysis adds a canonical perspective to the membrane paradigm. In this paradigm the null constraints have previously been read as conservation laws  \cite{damour1979quelques, Price:1986yy, Parikh:1997ma}, and linked to conservation laws at null infinity in \cite{Penna:2015gza}, but their canonical meaning is not usually stressed. 

The null edge modes of gravity are relevant also to the thermodynamics of black hole and other horizons, to which our analysis also applies. This relates to work done on isolated horizons \cite{Ashtekar:2000hw}. 

Our central results are as follows: The Damour \cite{damour1979quelques} and Raychaudhuri equations, which are the null constraints and part of the Einstein equations, are interpreted as conservation equations on a null surface $B$ for  a \emph{boundary current} $J_\xi$ associated with an arbitrary vector field $\xi \parallel B$. The boundary current is different from the Komar superpotential, and is given entirely in terms of data associated with the geometry of $B$. More precisely, we show that the null constraints can be understood as expressing the divergence of the boundary current as the sum of the matter energy-momentum flux  plus the \emph{gravitational energy-momentum flux} $F_\xi$. The null constraints are summarized as one equation on densities on $B$:
\begin{align}\label{intro:cons}\boxed{
	\rd J_\xi = T_{L \xi} \eb + F_\xi.}
\end{align}
On the RHS, $T_{L\xi}$ is the matter energy-momentum tensor contracted with $\xi$ and the null normal $L$, and $\eb$ is a volume element on $B$. The gravitational flux $F_\xi$ is of the canonical form
\begin{align}\label{eq:intro_flux}
	F_\xi = \sum_i P_i \LL_\xi Q_i.
\end{align}
Here, $(P, Q)$ are the canonical bulk pairs of gravity on a null surface. We show in two different ways that the set of canonical pairs $(P_i, Q_i)$ consists of a spin-2 pair $(\tfrac12 \sigma^{AB} \eb, \gamma_{AB})$ composed of the densitized shear and the conformal metric on spatial cross-sections of $B$; a spin-1 pair $(- \omega_a \eb, L^a)$ made of the \emph{twist} $\omega_a$ and the null generator $L^a$ of $B$; and a spin-0 pair $(\eb, \mu)$ that contains  the null volume density $\eb$ on $B$ and the \emph{spin-0 momentum} $\mu$. The spin-0 momentum is the linear combination
\begin{align}
	\mu = \kappa + \tfrac{D-3}{D-2} \theta
\end{align}
of the null acceleration $\kappa$ and expansion $\theta$, where $D$ is the dimension of spacetime.  $\mu$ plays the role of a gravitational pressure for the boundary spheres.

The \emph{field space Lie derivative} $\LL_\xi$ in (\ref{eq:intro_flux}) describes the action of infinitesimal diffeomorphisms $\xi\parallel B$ on the data $(P, Q)$ which are viewed as functionals of the full metric. This action is non-trivial and is given explicitly in the bulk of the paper.
The flux term (\ref{eq:intro_flux}) can be understood as arising from the \emph{intrinsic symplectic potential} on $B$,
\begin{align}
	\Theta[g_{ab}, \delta g_{ab}] = \tfrac12 \eb \sigma^{AB} \delta \gamma_{AB} -  \eb \omega_a\delta L^a  + \eb \delta \mu.
\end{align}
The flux is simply given by the canonical expression $F_\xi = \Theta[g_{ab}, \LL_\xi g_{ab}]$, which is of the same form as the canonical matter energy-momentum tensor. The intrinsic symplectic potential that one need to use in order to achieve these results \emph{differs} from the standard choice for the gravity symplectic potential by a (codimension two) corner term. This boundary term is part of the standard ambiguity \cite{Jacobson:1993vj} that appears in the definition of the symplectic potential, and is fixed here by the demand that the symplectic potential depend only on the intrinsic and extrinsic geometry of $B$.
Correspondingly, the boundary current in our analysis differs from the standard Komar charges in two ways: it does not require the vector field $\xi$ to be extended outside of the null surface, and it contains only the intrinsic and extrinsic geometry of the null surface and no additional data.

As an important result that highlights the canonical meaning of the constraints,  we show that the boundary current $J_\xi$ in (\ref{intro:cons}) coincides with the Noether charge form of the {\it intrinsic} symplectic potential.  
For a vector field $v$ parallel to fixed spatial cross-sections $S$ of $B$ the boundary current reads
\begin{align}
	J_v = \big((v^b \omega_b) L^a + v^b \sigma_b{}^a \big)\iota_a \eb,
\end{align}
where $\iota$ is the contraction of vectors with forms. It contains the \emph{momentum aspect} $(v^b \omega_b) \epsilon_S$ where $\epsilon_S=\iota_L \eb$ is the area form on $S$, which, when integrated on a codimension $2$ sphere  describes the amount of super-momentum within a region. The integrated momentum aspect is also the Hamiltonian generator of the infinitesimal diffeomorphism $v$. The boundary current also contains a spatial momentum current $(v^b \sigma_b{}^a) \iota_a \eb$. 

For a null vector field $fL^a$ tangential to $B$, the boundary current reads
\begin{align}
	J_{fL} = ( f (\mu-\theta) + L[f]) \es.
\end{align}
It can be identified with the \emph{energy aspect}, and contains no spatial current.
As in  thermodynamics, the notion of which  gravitational energy is appropriate depends on which variables are chosen to be controlled on the null surface. The energy aspect we have given is analogous to the enthalpy, it corresponds to controlling the spin-0 momentum $\mu$, which plays the role of boundary pressure. The choice of $\mu$ is shown to be entirely determined by a choice of  ``clock'' along $B$, i.e., by the normalization of the null normal. We show that under boundary conditions which fix $\mu$ and the shear $\sigma^{ab}$ on the boundaries of $B$, and when the intrinsic symplectic form is used, the energy aspect $J_{fL}$ is the Hamiltonian generator of infinitesimal diffeomorphisms along $fL$.

The plan of the paper is as follows: Section \ref{sec:geom} defines the intrinsic and extrinsic geometry of the null surface and gives the action of the field space Lie derivative. Section \ref{sec:eeq_as_conservation} rearranges the Damour and Raychaudhuri equations as a canonical conservation equation, and derives the boundary current $J_\xi$ and the flux terms $F_\xi$. Section \ref{sec:from_sp} addresses the rationale and consequences of modifying the symplectic potential away from the standard one, and derives the canonical conservation equation starting from the intrinsic symplectic potential. Section \ref{sec:hams} addresses the question how the boundary current $J_\xi$ is related to the Hamiltonian generators of the infinitesimal diffeomorphism $\xi$. Technical  manipulations have been relegated to appendices.

\section{Geometry}\label{sec:geom}
This section introduces the coordinates, the parametrization of the metric, and the intrinsic and extrinsic geometry of a null surface used throughout our analysis.
\subsection{Spacetime Metric and Intrinsic Geometry of a Null Surface}

We work in $D$ dimensions, and use coordinate fields $X^a$ on a region of the spacetime $M$.
\be
X^a(x)= (\phi^0(x), \phi^1(x), \sigma^A(x)).
\ee 
The fields $\phi^i, i \in \{0,1\}$, foliate the region into spacelike codimension $2$ spheres $S_{u,r}$ given by the level surfaces $(\phi^0,\phi^1)=(u,r)$. The spheres are coordinatized by $\sigma^A$, with $A \in \{2, ..., D-1\}$. The level surfaces of $\phi^0$ are taken to be spacelike or null with $\phi^0$ increasing towards the future.
The level surface of $\phi^1$ are taken to be timelike or null.

We assume that a null hypersurface $B$ of cylinder topology is situated at $\phi^1 = 0$. On $B$, we use coordinates $(\phi^0, \sigma^A)$, such that the embedding $\jmath_B: B\hookrightarrow M$ becomes $\jmath_B: (\phi^0, \sigma^A) \mapsto (\phi^0, \phi^1 = 0, \sigma^A)$. The surface $B$ is foliated into spacelike $(D-2)$-spheres $S_{u}$ defined by the condition  $\phi^0=u$.  Finally, the surface $B$ has past and future boundaries, which are assumed to be at constant $\phi^0$.

Using the coordinate fields, a general spacetime metric can be parametrized in terms of a normal metric $H_{ij}$, a  normal  connection $A_i^A$ valued into the sphere tangent bundle and a $(D-2)$-dimensional metric $q_{AB}$ as
\begin{align}
	\rd s^2 =  H_{ij} \rd \phi^i \rd \phi^j + q_{AB} (\rd \sigma^A - A^A_i \rd \phi^i) (\rd \sigma^B - A^B_j \rd \phi^j).
\end{align}
This is the codimension two generalization of the familiar ADM lapse-shift form, with the normal metric generalizing the lapse function and the normal connection generalizing the shift vector. The measures on spacetime $M$ and the spheres $S$ are related as $\sqrt{|g|} = \sqrt{|H|}\sqrt q$. We parametrize the normal metric in terms of 3 scalar parameters $(h,\beta,\bar\beta)$ as
\begin{align} \label{eq:H_param}
	H_{ij} = \frac{e^{h}}{1+\beta\bb}
	\begin{pmatrix}
		-2 \beta & 1-\beta\bb\\
		1-\beta\bb & 2 \bb
	\end{pmatrix},\qquad H_{ij} \rd \phi^i \rd \phi^j
	= \frac{2e^{h}}{1+\beta \bar{\beta}} ( \rd \phi^0 +\bar{\beta} \rd \phi^1) (\rd \phi^1 - \beta \rd \phi^0).
\end{align}
Its determinant is $|\det(H)| = e^{2h}$ and the parameter $\beta$ vanishes on $B$ since $B$ is null. We restrict $\beta, \bar{\beta}\geq 0$  to ensure that the hypersurfaces $\phi^0=u$ are spacelike or null and that the hypersurfaces $\phi^1=r$ are timelike or null. 

To capture the null geometry of the sphere foliation, let us introduce on $M$ the null vector field $L= L^a\pa_a$  and the null one-form field $\Lb=\Lb_a \rd x^a$ given by
\begin{align}\label{dualpair1}
	L:= D_0 + \beta D_1, \qquad \bar{L} := \frac{1}{1+\beta\bar\beta} (\rd \phi^0 +\bar\beta\rd \phi^1)
	,\qquad D_i := \frac{\p}{\p \phi^i} + A_i^A \frac{\p}{\p \sigma^A},
\end{align}
where we defined the vectors $D_i$, which are normal to the spheres $S_{u,r}$. The vector $L$ is normal to the spheres, future pointing and normalized by the condition\footnote{$\xi[f]$ denotes the directional derivative of a function $f$ along a vector field $\xi$.} $L[\phi^0] = 1$. The form $\Lb$ is dual to $L$ and normalized by $\iota_L \Lb=1$ with $\iota$ denoting the vector  contraction.
The vector $L$ is parallel to $B$, and can thus be viewed as intrinsic to $B$:
\begin{align}
	L \overset{B}{=} D_0.
\end{align}
Also note that on $B$, the form $\Lb$ is exact, i.e., a total differential: $\jmath_B^* \Lb = \rd \phi^0$.

Using the metric, we can construct the dual pair $(g(L),g^{-1}(\bar{L}))= (L_a\rd x^a, \bar{L}^a \pa_a)$, explicitly
\be \label{eq:dualpair}
{g}(L)= e^{h}(\rd\phi^1 -\beta\rd\phi^0),\qquad
{g}^{-1}(\bar{L})=\frac{e^{-h}}{1+\beta\bar\beta} (D_1 - \bar\beta D_0).
\ee
The vector $g^{-1}(\Lb)$ is normal to the spheres $S$, while 
$g(L)$ is the null normal of $B$. Neither $g(L)$ nor $g^{-1}(\Lb)$ are intrinsic to $B$, so they will not appear in the covariant  expressions we will introduce. The vector $g^{-1}(\Lb)$ is past pointing and null. The full space-time metric can be decomposed as\footnote{Strictly speaking, we have $q_{ab} = q_{AB} e_a{}^A e_b{}^B$ with the projectors $e_a{}^A = \frac{\p \sigma^A}{\p x^a}\vert_{\phi^i = \const} - A^A_a$. We will suppress projectors and also use the same indices $(a, b, ...)$ on $B$ and $M$ for compactness of notation.}
\begin{align}
	g_{ab} = q_{ab} + L_a \Lb_b + \Lb_a L_b.
\end{align}

Since the conformal structure and the determinant on the spheres play separate roles, we parametrize the sphere metric $q_{AB}$ in terms of a conformal factor $\varphi$ and a conformal metric $\gamma$ of unit determinant:
\be
q_{AB} = e^{2\varphi} \gamma_{AB},\qquad \det(\gamma)=1. 
\ee
The conformal factor determines the luminosity distance  $R=e^\varphi$.
The measure $\sqrt q$ on the sphere $S$ and the spacetime measure $\sqrt{|g|}$ are then given by 
\begin{align}
	\sqrt{q}=  e^{(D-2) \varphi}, \qquad \sqrt{g}=e^h e^{(D-2) \varphi}.
\end{align}

Let us now turn to the induced geometry of $B$. The  metric on $B$ is
given by the  pullback $\jmath_B^*(\rd s^2)$ of the full metric along the inclusion $\jmath_B:B \hookrightarrow M$. It is degenerate and reads
\begin{align}
	\rd s^2_B ={}& e^{2\varphi} \gamma_{AB} (\rd \sigma^A - A^A_0 \rd \phi^0)(\rd \sigma^B - A^B_0 \rd\phi^0).
\end{align}
The induced geometry on $B$ is thus determined by $(\varphi, \gamma_{AB}, A_0^A)$, or equivalently by $(\varphi, \gamma_{AB}, L^a)$. 

Since the metric on $B$ is degenerate, it does not have a preferred volume $(D-1)$-form. However, there is an covariant area $(D-2)$-form $\es$. Let $\epsilon$ be the volume form\footnote{In our coordinates it is explicitly given by $\epsilon =e^{h+(D-2)\varphi}  \rd \phi^0 \wedge \rd \phi^1 \wedge \rd^{D-2} \sigma$.} on $M$. The area form on $B$ is given by\footnote{In our coordinates it reads 
	\begin{align}
		\es =  e^{(D-2) \varphi}\tfrac{1}{(D-2)!} \epsilon_{A_3 ... A_D} (\rd \sigma^{A_3} - A_0^{A_3} \rd \phi^0) \wedge ... \wedge (\rd \sigma^{A_D} - A_0^{A_D} \rd \phi^0).
	\end{align}
	It is invariant under the redefinitions $L\to e^\alpha L$ and $g^{-1}(\Lb)\to e^{-\alpha}(g^{-1}(\Lb)+v)$ where $v$ is tangent to $B$, and covariant under diffeomorphisms of $B$.
} 
\be
\es :=\jmath_B^*(  \iota_{g^{-1}(\Lb)} \iota_L \epsilon)
\ee 
where again $\iota$ is the contraction of  vectors with forms. The  pullback of $\es$ to any cross-section $S_u$ of $B$ coincides with the induced volume form $\rd S: =\sqrt{q} \rd^{(D-2)}\sigma$ on the cross-section, i.e., we have $ i_S^* \es = \rd S$ with the inclusion $i_S: S \hookrightarrow B$. The area form is orthogonal to the null directions, we have $\iota_L \es = 0$.

We also introduce a volume $(D-1)$-form
on $B$ given by
\begin{align}
	\eb := \rd \phi^0\wedge \es = \rd \phi^0 \wedge \rd S.
\end{align}
It can also be defined as $\eb = - \jmath_B^* (\iota_{g^{-1}(\Lb)} \epsilon)$ and is related to $\es$ as $\iota_L \eb = \es$. A $(D-1)$-form $\iota_\xi \epsilon$ then pulls back to $B$ as
\begin{align}
 \jmath_B^* (\iota_\xi \epsilon) = - L_a \xi^a \eb.
\end{align}
The null vector $L^a$ ruling $B$ is subject to the normalisation condition $L[\phi^0]=1$, and thus depends on a choice of ``time'' foliation of $B$. This will be reflected in its transformation under diffeomorphisms (see section \ref{ssec:diffeos}). Similarly, the form $\eb$ depends on the choice of $\phi^0$. However, their combination $L^a \eb$ does not depend on the choice of $\phi^0$ and transforms covariantly under diffeomorphisms of $B$. Since at $B$, $L$ is parallel to $B$, we can view $L^a \eb$ as an object intrinsic to $B$.
		
To summarize, the intrinsic geometry of $B$ is captured by the conformal $(D-2)$ metric $\gamma_{AB}$, the conformal factor $\varphi$ and the null direction $L^a$. This data determines a covariant vector valued $(D-1)$-form $ L^a \eb$, and a covariant area $(D-2)$-form $\es$.

\subsection{Partial Gauge Fixing}

In the construction of the symplectic potential done in \cite{Hopfmuller:2016scf} we have shown that the metric parameters
$\bar{\beta}$ and $A_1^A$ do not enter the symplectic potential if one restricts to variations that leave $B$ null. They can therefore safely be gauge fixed at $B$ without loosing any degrees of freedom. We stress that while the partial gauge fixing is useful for physical interpretation and for cleaning up some coordinate expressions, our analysis does not rely on this or any gauge fixing. We introduce the partial gauge fixing
\be
\bar\beta=0,\qquad A_1^A=0 
\ee 
and to agree with usage we denote $A_0^A $ in this gauge by the tangential vector $U^A$. In this gauge $D_1=\pa_1$ and  the radial vector $\pa_1$ is null and geodesic. 
Lines of constant $\phi^0$ and constant $\sigma^A$ are light rays, and the radial field $\phi^1$ is a parameter along them. This parameter is affine iff $\p_1 h = 0$. 

This gauge is similar to the Bondi  gauge 
which contains the additional gauge condition that the radial coordinate measures the size of the spheres, that is in the Bondi gauge we also demand that 
$\phi^1= e^{\pm\varphi}$, with $+$ for advanced and $-$ for retarded time \cite{Madler:2016xju}. 
Another condition commonly imposed by Penrose \cite{Penrose:1986ca} is to demand that $\phi^1$ is an affine parameter of the transverse light rays. The Penrose gauge therefore imposes that $\pa_1 h=0$.
 We will neither impose Bondi nor Penrose gauge in the following since none are preferred from the point of view of the canonical analysis, it will be handy to keep this freedom open.

In this gauge the full spacetime metric is parametrized by 3 scalars: $(h,\varphi, \beta)$, one vector $U^A$ on the sphere, and the conformal metric $\gamma_{AB}$. $h$ and $\varphi$ control the local scale of the normal geometry and tangential geometry respectively, and $\beta$ measures how much the surfaces $\phi^1 = \const$ deviate from being null. In this gauge the metric  reads
\begin{align}\label{eq:gf_metric}
\rd s^2 ={}& {2e^{h}} \rd \phi^0  (\rd \phi^1 - \beta \rd \phi^0)
+ e^{2\varphi} \gamma_{AB}  (\rd \sigma^A - U^A \rd \phi^0) (\rd \sigma^B - U^B \rd \phi^0).
\end{align}
The null vector $L$ and null form $\Lb$ can be expressed in terms of the coordinate derivatives $ D_0 := \frac{\p}{\p \phi^0} + U^A \frac{\p}{\p \sigma^A}$ and $\p_1$ as 
\begin{align}\label{dualpair2}
	L = D_0 + \beta \p_1\overset{B}{=}D_0, \qquad 
	\bar{L}  = \rd \phi^0.
\end{align}
In this gauge, $\bar{L} =  \rd \phi^0$ is thus an exact form not only on $B$ but on all of $M$. It is clear that the pair $(L,\bar{L})$ modulo rescaling 
$(e^a L, e^{-a}\bar{L})$ is intrinsic\footnote{Unlike 
the pairs $(g(L),g^{-1}(\Lb))$:
\be
{g}(L)=  e^h(\rd \phi^1-\beta \rd \phi^0)\overset{B}{=}e^h \rd \phi^1, \qquad {g}^{-1}(\bar{L})=
\pa_1,
\ee
which depends on the choice of transverse coordinate. Most of our equations involve $L^a$ and $\Lb_a$, but not $L_a$ and $\Lb^a$. } to the geometry of $B$.

The inverse metric in the partial gauge is given by 
\be
g^{ab}\pa_a \pa_b 
= 2e^{-h} (D_0+ \beta \pa_1)   \pa_1 + q^{AB} \pa_A \pa_B.
\ee
The null gauge therefore corresponds to the  conditions $g^{00}=0$, $g^{0A} = 0$.
Diffeomorphisms that preserve this gauge are therefore given by vectors that satisfy the conditions
\be
\pa_1 \xi^0=0, \qquad \pa_1 \xi^A = - e^{h}q^{AB} \pa_B \xi^0.
\ee

\subsection{Extrinsic Geometry of a Null Surface} \label{ssec:extrinsic}
 As we have seen the intrinsic geometry of $B$ is encoded into the data $(\gamma_{AB}, U^A, \varphi)$. As we will see,
the canonical momenta associated to this triple can be constructed in terms of the extrinsic geometry elements of the null surface $B$ as embedded in  $M$. This subsection therefore introduces these pieces of extrinsic geometry. Besides their role as momenta, they also appear as part of the fluxes and as charges.

In the following, we denote as $\nabla_a$ the covariant derivative of $g_{ab}$ on $M$, as $\rd_A$ the covariant derivative of $q_{AB}$ on $S$, and as $\rd$ without index the exterior derivative on $B$ and on $M$. $\L$ is the Lie derivative, and $\xi[f]$ the directional derivative along a vector field $\xi$ of a function $f$. The extrinsic geometry of $B$ is encoded into  the following:
\begin{itemize}
	\item The conformal shear is given by the Lie derivative along $L$ of the conformal metric: $ \sigma^{AB} :=\gamma^{AA'} \gamma^{BB'}\frac12 {\cal L}_L \gamma_{A'B'}$ where $\gamma^{AB}$ is the inverse of $\gamma_{AB}$. It is automatically trace free, and can be defined from the trace free part of the extrinsic curvature\footnote{We denote the trace free components of a tensor as $		\theta^{<AB>}= \theta^{AB}-\tfrac{q^{AB}}{(D-2)} q^{CD}\theta_{CD}$} as 
	$\sigma^{AB} =  e^{2\varphi} \theta^{<AB>}$ with $\theta^{AB}= q^{Aa} q^{Bb}\nabla_a L_b$. It gives the canonical  momentum conjugate to the conformal metric $\gamma_{AB}$.
	\item The twist field $\omega_A := q_A{}^a (\bar{L}_b \nabla_a L^b)$. It gives the momentum conjugate to $U^A$, and the charge for infinitesimal diffeomorphisms of the cross-sections $S$.
	\item The expansion $\theta = \frac12 q^{AB} {\cal L}_L q_{AB}$, 
	which enters the conservation of the canonical two form via $\rd \epsilon_S = \theta \epsilon_B$. It plays a central role in the Raychaudhuri equation.
	\item The surface gravity $\kappa$, defined on $B$ as the null acceleration $ \nabla_L L^a \overset{B}{=} \kappa L^a$.
\end{itemize}

The elements of extrinsic geometry appear naturally in the comparison between two different ways of transporting a vector field $\xi$ on $B$: the Lie transport $\L_L \xi$, which is purely intrinsic to $B$, and the parallel transport $\nabla_L \xi$, which through the Christoffel symbols contains information about the metric components transverse to $B$. Let us decompose the vector field $\xi\in \Gamma(TB)$ as 
\begin{align}
\xi^a = f L^a + v^a,
\end{align} 
with $v \parallel S$. The difference between the parallel and Lie transport along $L$ of a vector $\xi$ tangent to $B$  is given by 
\be\label{eq:nabla_L_xi_into_B}
\nabla_L \xi^a - [L,\xi]^a = \nabla_\xi L^a=   v^b (  \omega_b L^a+ \sigma_b{}^a +\tfrac{\delta_b^a}{D-2}\theta ) + f \kappa L^a.
\ee
This difference is encoded into the so-called Weingarten map $\nabla_\xi L^a$ , see e.g. \cite{Gourgoulhon:2005ng}.
The elements of the extrinsic geometry enter the expansion of  the Weingarten map in terms of ``spin two'' components $\sigma_{AB}$, ``spin one'' components $\omega_A$ and ``spin zero'' components $(\kappa,\theta)$. 
We can express each component in terms of the metric coefficients:

\textbf{Spin 2:} The shear in terms of our parametrization of the metric becomes
\bea
\sigma_{AB} &=& \tfrac12 D_0 \gamma_{AB}  + \tfrac12 (\gamma_{BB'}\pa_{A} U^{B'} + \gamma_{AA'}\pa_{B} U^{A'})  - \tfrac1{D-2} \gamma_{AB} \pa_C U^C,\cr
&=& \tfrac12 \pa_0\gamma_{AB} + e^{- 2 \varphi} \rd_{<A} U_{B>}.
\eea
Note that $e^{-2 \varphi} \rd_{<A} U_{B>}$ is independent of the conformal factor $\varphi$, so $\sigma_{AB}$ depends just on $\gamma_{AB}$ and $U^A$. Although the shear is part of the extrinsic geometry of $B$, it is determined by the intrinsic geometry. If we interpreted $U^A$ as the velocity field of a fluid on $B$, the term $e^{-2 \varphi} \rd_{<A} U_{B>} $ is naturally interpreted as the rate of strain tensor. It is complemented by the time derivative of the metric in the case where the metric is explicitly time dependent, which is not usually the case in fluid dynamics. Also note that both $\gamma_{AB}$ and $\sigma^{AB}$ are invariant under conformal rescalings of the metric.

\textbf{Spin 1:} 
An alternate definition is available for the twist field $\omega_A$: we have
\begin{align}\label{omegaev}
 \omega_A = \bar\eta_A:= - q_A{}^b \nabla_L \bar{L}_b,
\end{align}   
which represents the parallel transport of the dual one-form $\bar{L}_a$ along $L$. To prove the identity, first note $\omega_A = q_A{}^a \Lb_b \nabla_a L^b = - q_A{}^a L^b \nabla_a \Lb_b$, where we integrated by parts using $L^a \Lb_a = 1$. Now use that the pullback of $\Lb_a$ onto $B$ is just $\rd \phi^0$, i.e., it is exact. Hence, when contracted with two vectors parallel to $B$, we have $\nabla_a \Lb_b = \nabla_b \Lb_a$. We thus have $\omega_A = - q_A{}^a L^b \nabla_a \Lb_b = - q_A{}^a L^b \nabla_b \Lb_a$ which proves the identity (\ref{omegaev}).

The coordinate expression for the twist $\omega_A$ is derived in the appendix of \cite{Hopfmuller:2016scf} and reads $\omega_A=\frac12 (\pa_A h - e^{-h}q_{AB}[D_0, D_1]^B)$, which using the partial gauge fixing becomes
\begin{align}
\omega_A= \frac12( \pa_A h + e^{-h} q_{AB} \pa_1 U^B).
\end{align}
Under conformal rescaling of the metric $g \rightarrow e^{2 \alpha} g$, the size of the normal geometry transforms as $h \rightarrow h + 2 \alpha$ while $U$ does not change. The twist then transforms by a total derivative, $\omega_A \rightarrow \omega_A + \p_A \alpha$, and its curvature $\rd_{[A} \omega_{B]}$ is conformally invariant.

\textbf{Spin 0:} The spin-0 sector is especially interesting, since it carries information about mass and energy. A wide variety of different linear combinations of the spin-0 variables $\kappa$ and $\theta$ appear in the literature. The conformally invariant combination is $\kappa - \tfrac{2}{D-2}\theta$, it is  constant on conformal Killing horizons \cite{DeLorenzo:2017tgx} (recall that $D$ is the dimension of spacetime $M$). 
The combination $\kappa - \tfrac{1}{D-2} \theta$ will appear in our charges and Hamiltonians. It also features in the null Raychaudhuri equation written as
\begin{align}
	G_{LL} = {}& - L[\theta] + \left(\kappa - \tfrac{1}{D-2} \theta\right)\theta - \sigma_{A}{}^B \sigma_B{}^A,
\end{align}
so if one sets that combination zero and knows $\sigma_A{}^B$, the equation can straightforwardly integrated for $\theta$ (such as in \cite{0264-9381-10-4-012}). The combination $\kappa + \theta$ is obtained as $\kappa +\theta = \nabla_a L^a$ and has been suggested as the null analogue of the Gibbons-Hawking-York term \cite{Parattu:2016trq}.
However the combination that is of crucial interest for us is the combination
\begin{align}\label{eq:def_mu}
	\mu := \kappa + \tfrac{D-3}{D-2}\theta
\end{align}
which we call the \emph{spin-0 momentum}. It enters our analysis as the canonical variable conjugate to the conformal factor $\varphi$, and naturally appears in the densitized Raychaudhuri and Damour equations as we will see.
This combination appeared in dimension $4$ in the canonical analysis of Torre \cite{Torre:1985rw} and of Epp \cite{Epp:1995uc}, see also \cite{DePaoli:2017sar} for its interpretation in the first order formalism. It combines the pressure and bulk viscosity terms from the membrane paradigm \cite{Price:1986yy}.

In terms of our parametrization of the metric, the expansion becomes
\be
\theta = {(D-2)}   D_0 \varphi + \pa_A U^A.
\ee
Even though it is part of the extrinsic geometry of $S$, it is determined by the intrinsic geometry of $B$. That can also be seen noting that the divergence of the area form on $B$ is $\rd \es = \theta \eb$. More generally, we have for any function $g$ on $B$:
\begin{align}
\rd (g \es) = (L[g] + g \theta) \eb.
\end{align}
The coordinate expression for the acceleration is derived in \cite{Hopfmuller:2016scf} and reads
\be\label{ka}
\kappa  = (D_0 + \beta D_1) h + D_1 \beta \overset{B}{=} D_0 h + \pa_1 \beta.
\ee 
For the spin-0 momentum, we thus get
\begin{equation}
	\mu \overset{B}=  D_0(h + (D-3) \varphi) + D_1 \beta +\tfrac{D-3}{D-2} \pa_A U^A.
\end{equation}
The coefficients $\kappa$ and $\theta$ are not invariant under local rescalings of the 
metric. Under a change $g_{ab} \to e^{2\alpha} g_{ab}$, holding $L^a$ and $\Lb_a$ fixed, we have on $B$
\be
(\kappa, \theta ) \to  (\kappa + 2 D_0 \alpha, \theta + (D-2) D_0\alpha ).
\ee

To summarize, we have decomposed the extrinsic geometry of the null surface $B$ as embedded in spacetime $M$ into the shear $\sigma^{AB}$, the twist $\omega_A$, the expansion $\theta$ and the surface gravity $\kappa$. We have defined the spin-0 momentum $\mu = \kappa + \tfrac{D-3}{D-2}\theta$.

\subsection{Transformations of Intrinsic and Extrinsic Geometry under Diffeomorphisms}\label{ssec:diffeos}
We now turn to the transformations under infinitesimal diffeomorphisms of the pieces of intrinsic and extrinsic geometry of the null surface $B$, which we will need to understand the conservation laws.
The expressions we have introduced make reference to the coordinate fields, especially to the ``time'' variable $\phi^0$ in the normalization of $L$. Under diffeomorphisms, we thus cannot expect our variables to transform covariantly, in the sense illustrated by the following example:

The null vector $L \in \Gamma(TB)$ can be Lie-derived along a vector field $\xi \in \Gamma(TB)$ in two ways: On the one hand, we have the standard Lie derivative on $B$ which acts on $L^a$ as a vector:
\begin{align}\label{eq:L_L}
	(\L_\xi L)^a = \xi^b \p_b L^a - L^b \p_b \xi^a.
\end{align}
On the other hand, we can view every component of $L^a= L^a(g_{bc})$ as a function of components of the metric. Since we know how metric components transform, that fixes the transformation of the components of $L^a$. We call this procedure the \emph{field space Lie derivative} and denote it as $\LL_\xi$.  This 
field space Lie derivative is essential to us since it is the one that enters the canonical analysis. 
In practice the field space Lie derivative of a functional $F(g_{bc})$ is simply given by $\LL_\xi(F(g_{bc}))= \frac{\pa F(g_{bc})}{\pa g_{ab}}\LL_\xi g_{ab}$, and we use that $\LL_\xi g_{ab}= {\cal L}_\xi g_{ab}$.  Concretely, on $B$ the components of $L$ are $L^0 = 1, L^A = U^A$, which we can summarize as
\begin{align}
	L^a(g_{bc}) \overset{B}{=} \frac{g^{1a}}{g^{10}}.
\end{align}
Using the Leibniz rule, we get
\begin{align}
	\LL_\xi L^a = \frac{1}{g^{10}} \LL_\xi g^{1a} - \frac{g^{1a}}{(g^{10})^2} \LL_\xi g^{10}.
\end{align}
Now use that the components of the inverse metric transform by the spacetime Lie derivative:
\begin{align}
\LL_\xi (g^{ab}) = (\L_\xi (g^{-1}))^{ab} = \xi^c \p_c g^{ab} - g^{ac} \p_c \xi^b - g^{cb} \p_c \xi^a.
\end{align} 
Parametrizing as before $\xi^a = f L^a + v^a$ with $v \parallel S$, we get after a short calculation
\begin{align}
	\LL_\xi L^a = [v, L]^a,
\end{align}
which does not coincide with (\ref{eq:L_L}).

To formalize the notion of a field space Lie derivative, it is useful to think of field space, i.e., the space of metrics, as a differentiable manifold, and of every component of the intrinsic and extrinsic geometry tensors as a function on field space. Introducing the exterior derivative on field space $\delta$, which we will call the \emph{variation}, and the contraction $I_\xi$ on field space allows us to write
\begin{align}
\LL_\xi F(g) = I_\xi \delta F(g)
\end{align} 
for any field space function\footnote{We will later extend $\LL_\xi$ to field space forms via the Cartan formula $\LL_\xi = I_\xi \delta + \delta I_\xi$.} $F(g)$. For example, we have $\LL_\xi L^a = I_\xi \delta L^a$. Now note that on $B$ we have 
\begin{align}
 \delta L^a = - q^{ab} L^c \delta g_{bc},
\end{align} 
which may be checked directly using the parametrization (\ref{eq:gf_metric}). Since the $\phi^0$-component of $L$ is fixed, it does not vary, and $\delta L^a = \delta U^a$ is parallel to the cross-sections $S$. Using also $\LL_\xi (g_{ab}) = (\L_\xi g)_{ab} = \nabla_a \xi_b + \nabla_b \xi_a$, we get that $\LL_\xi L^a = - q^{ab} L^c (\nabla_b \xi_c + \nabla_c \xi_b)$, which provides an alternative way of calculating field space Lie derivatives.

A useful bookkeeping device is the difference between the field space and spacetime Lie derivatives, which we call the \emph{anomaly} $\Delta$:
\begin{align}
	\Delta_\xi := \LL_\xi - \L_\xi.
\end{align}
For example, on $B$, $\Delta_\xi L^a = [v, L]^a - [\xi, L]^a = L[f] L^a$. We call a tensor \emph{covariant} if it has vanishing anomaly. The anomaly of the null vector $L$ can be understood as the source of all the anomalous diffeomorphism transformations. It stems from the normalization condition $L[\phi^0] = 1$, which introduces the field $\phi^0$ as background structure and thus breaks covariance. Since there is no preferred normalization of the null normal, some degree of non-covariance is unavoidable when dealing with null surfaces.

The covariance of the covariant derivative is encoded as
\begin{align}
	\Delta_\xi \nabla_a T = \nabla_a \Delta_\xi T
\end{align}
for any tensor $T$, which needn't be covariant. The identity may be checked explicitly using the standard identity $\delta \Gamma^a_{bc} = \frac12 (\nabla_b \delta g^a{}_c + \nabla_c \delta g^a{}_b - \nabla^a \delta g_{bc})$. We make use of this in appendix \ref{ap:diffeos} to derive the diffeomorphism transformations of extrinsic geometry.

The field space Lie derivatives of all the data we have defined so far are derived  in the appendix \ref{ap:diffeos}, and we summarize the relevant results now. As before, let $\xi^a = f L^a + v^a$ be a vector field on $B$, and let $v \parallel S$.
In addition to the transformation of $L$, we will have the following:
The conformal metric transforms as would be expected,
\begin{align}
	\LL_\xi \gamma_{AB} = 2 (f \sigma_{AB} + e^{-2 \varphi} \rd_{<A} v_{B>}),
\end{align}
where we recall that $\rd_A$ is the covariant derivative of $q_{AB} = e^{2\varphi} \gamma_{AB}$. Note that $e^{-2\varphi} \rd_{<A} v_{B>}$ is independent of $\varphi$, and that the RHS is trace free as expected of derivatives of a unimodular matrix. The conformal factor transforms as
\begin{align}
	\LL_\xi \varphi = \tfrac{1}{D-2} (f \theta + \rd_A v^A).
\end{align}
As argued earlier, the combination $L^a \eb$ is covariant, it transforms under diffeomorphisms of $B$ as
\begin{align}
	\LL_\xi (L^a \eb) = \L_\xi (L^a \eb) = \big([v,L]^a +  (\theta f + \rd_B v^B) L^a \big) \eb.
\end{align}
Here, the spacetime Lie derivative acts on $L^a \eb$ as a vector valued top form, i.e., $\L_\xi (L^a \eb) = [\xi, L]^a \eb + L^a \rd (\iota_\xi \eb)$.
The area $(D-2)$-form is covariant, and by contracting the previous we obtain
\begin{align}
	\LL_\xi \es = \L_\xi \es = \iota_\xi \rd \es + \rd \iota_\xi \es = \big([v, L]^a \iota_a \eb + (\theta f + \rd_B v^B)\big) \es.
\end{align}
Finally, we need the transformation of the spin-0 momentum, which is more subtle. Under finite rescalings of the null generators $L \rightarrow g L$ (or equivalently under redefinition of the coordinate $\phi^0$ with $\p \phi^0{} / \p \phi^0{}' = g$), the spin-0 momentum transforms as a connection and goes to
\begin{align}
	\mu \rightarrow \mu_g :={}& (L + \mu)[g].
\end{align}
The spin-0 momentum can thus be fixed to any value by controlling the ``clock'' $\phi^0$. We will make use of that fact in section \ref{sec:hams}. Infinitesimally, the transformation involves a second derivative of the vector field $\xi$, and reads:
\begin{align}
	\LL_\xi \mu = v[\mu] + L\big[(L + \mu)[f]\big].
\end{align}
Note the appearance of the differential operator $L+\mu$, which is a covariant derivative with respect to local rescaling of the null generators\footnote{Under $L \rightarrow e^\alpha L$ holding $f L$ fixed, we have $f \rightarrow e^{-\alpha} f$ and $\mu \rightarrow e^\alpha(\mu + L[\alpha])$, so $ (L + \mu)[f]$ is invariant.}.

Two remarks are in order: Firstly, the transformations of $(\varphi, \gamma, L, \mu, \omega_A, \sigma^{AB})$ only involve $\xi$ as a vector field on $B$, and do not depend on how (and if) it is extended to a vector field on $M$ (proof in appendix \ref{ap:diffeos}). This is far from obvious looking at the coordinate expressions given in section \ref{ssec:extrinsic}, and is an important and desirable feature of those variables. Secondly, note that $\mu, L, \eb$ and $\omega_A$ transform covariantly under diffeomorphisms $v$ parallel to the cross-sections $S$: Anomalies arise only for diffeomorphisms transverse to $S$.

We will also need the transformation of $h$, the logarithmic determinant of the metric in directions normal to $S$ (see \ref{eq:gf_metric}). It depends on the extension of $\xi$, and we parametrize an arbitrary extension as $\xi^a = fL^a + \bar f \Lb^a + v^a$ with $\bar f$ vanishing on $B$. Using $\delta h = L^a \Lb^b \delta g_{ab}$, we get
\begin{align}\label{eq:LL_xi_h}
	\LL_\xi h = (L + \kappa)[f] + (g^{-1}(\Lb) + \bar\kappa)[\bar f] + (\eta_A + \omega_A) v^A,
\end{align}
where $\eta_A = - q_A{}^a \nabla_{\Lb} L_a$ and $\bar\kappa = L_a \nabla_{\Lb} \Lb^a$. Note that $\eta$ and $\bar\kappa$ are not part of the extrinsic geometry of $B$, but rather part of the extrinsic geometry of $S$ as embedded in $M$.

To summarize, the tensors that make up the intrinsic and extrinsic geometry of $B$ can be Lie-derived in two ways: The spacetime Lie derivative views them as tensors and Lie-derives them according to their index structure, and the field space Lie derivative views them as functionals of the metric and derives them according to their metric dependence. The difference between the two prescriptions is the anomaly $\Delta_\xi$. We have given the field space Lie derivatives that we will need in the following, some more transformations are in the appendix \ref{ap:diffeos}.

\section{Einstein Equations as Conservation Equations}\label{sec:eeq_as_conservation}

Having completed the setup, 
let us turn to our central task of interpreting the Einstein constraint equations as 
conservation equations. 
We are looking for a conservation equation intrinsic to the null surface $B$ which is of  the form {``Divergence of current = gravitational flux + matter energy-momentum flux''}. Both the current and the gravitational flux will depend on a vector field $\xi$ which may be thought of as an observer. The conservation equation is an equation for $(D-1)$-forms, and can be integrated on portions of the null surface $B$.

The current on the LHS is the \emph{boundary current} 
$j_\xi^a$. It is a vector tangent to $B$ and we can associate with it a $(D-2)$-form  $J_\xi = \iota_{j_\xi} \eb$.
$J_\xi$  is a codimension one form on $B$ and the divergence of the current corresponds to  $\rd J_\xi$. 
In the following we will interchangeably use the denomination boundary current for $j_\xi$ or $J_\xi$ even if the later is the dual boundary form. The boundary current $j_\xi$ can be expanded in terms of a time component,
i.e., the component along $\phi^0$, and a component tangential to the sphere. Its time component may be thought of as the gravitational charge aspect, and the spatial components as the finite boundary analogue of soft currents.

The Einstein equations we consider are the null Raychaudhuri equation \cite{Raychaudhuri:1953yv} for $G_{LL}$
 and the Damour equation   \cite{damour1979quelques} for $q_a{}^bG_{Lb}$.
These are derived, for the reader's convenience, in appendix \ref{ap:eeq}.
This set of equations are the null analogue of the ADM momentum constraint equations. Since we are looking for a conservation law that can be integrated on the null surface $B$, we multiply them with the density $\eb$. The densitized expressions are
\bea\label{Ray1}
\boxed{\begin{aligned}[rcl]G_{LL}\, \eb &=&  -\L_L (\theta \eb) + (\mu \theta  - \sigma_b{}^a\sigma_a{}^b) \eb,\\
q_a{}^b G_{Lb}  \eb &=&    q_a{}^b \L_L(\omega_b \eb)- (\rd_a \mu  + \rd_b \sigma_a{}^b)\, \eb.\label{Damour1} \end{aligned}}
\eea
Note that densitizing with $\eb$ naturally leads to the appearance of the spin-0 momentum $\mu = \kappa + \tfrac{D-3}{D-2}\theta$ in both equations. 
Let us analyze them as conservation equations on $B$, first when contracted with a ``constant'' vector field, and then for a general vector field.

\subsection{Conservation Law for ``Constant'' Vector Fields}
To gain a first understanding, consider the Raychaudhuri and Damour equations smeared with a vector field $\xi =f L+ v$ parallel to $B$ which is Lie dragged along $L$, i.e.,
\begin{align}
	[L, \xi] = 0.
\end{align} 
This simplifying assumption, which is usually used at null infinity, means that $\xi$ is ``constant in time'' and implies $L[f] = 0$ and $[L, v] = 0$. It is sensitive to the choice of normalization of $L$, i.e., to a choice of clock.
Setting $G_{ab}=T_{ab}$ (in units where $8\pi G=1$) and contracting with $\xi$, we can rewrite our two equations as
\bea\label{eq:raych_cst}
- \L_L(f \theta \eb)  = f \left[T_{LL} - \mu \theta + \sigma_b{}^a\sigma_a{}^b  \right] \eb,\\
\label{eq:dns_cst} \L_L (v^a \omega_a \eb) =   v^a\left[T_{a L} + (\rd_a\mu- \rd_b \sigma_a{}^b ) \right]  \eb,
\eea
where we used $\L_L (g \es) = (L[g] + g \theta) \eb$ for any function $g$. Since $\L_L (g \eb) = \rd (g \es)$, the LHSs of both equations are total derivatives.

Written in this manner the Raychaudhuri equation (\ref{eq:raych_cst}) can be understood as  a conservation equation for an energy $E_f:=-\int_S f \theta \rd S$.
Indeed, by integrating the Raychaudhuri equation on a portion of $B$ delimited by $S_i$ and $S_f$, one gets the balance equation 
$\Delta E_f = \int_{S_i}^{S_f} f (T_{LL} +T_{LL}^G) \eb $, which expresses that the change in energy $E_f$ is due to exchange of material and gravitational energy with the exterior. 
This  allows us to identify the gravitational energy momentum tensor
\be \label{eq:cst_em_tensor}
T^G_{LL} := (\sigma_b{}^a\sigma_a{}^b - \mu \theta),
\ee
which appears alongside the matter energy-momentum tensor and measures the amount of gravitational energy that leaves the region enclosed by $S$ per unit time and unit area, according to the observer $\xi$. 
Part of the gravitational energy is carried out by the gravitational waves or spin 2 components $\sigma_b{}^a\sigma_a{}^b$, but another part is carried out by the spin zero component and measures the work done by the rescaling of the surface through the term $-\mu\theta$.
This naturally leads to the interpretation of $\mu$ as a boundary pressure term.

In the Damour equation (\ref{eq:dns_cst}), $P_v:= \int_S (v\cdot \omega)\rd S$ is interpreted as the super-momentum enclosed by the region $S$. 
We can identify a gravitational momentum flux 
$T^G_{\bm v L} $ given by 
\be
T^G_{\bm v L} = v^a (\rd_a\mu -  \rd_b \sigma_a{}^b).
\ee
This expression confirms the interpretation of $\mu$ as a pressure term, while the shear $\sigma$ appears as a viscous stress component.
Integrating the Damour equation then gives the balance equation 
$\Delta  P_v = \int_{S}^{S'} (T_{Lv} +T_{Lv}^G) \eb $.

\subsection{The Boundary Current and its Conservation}
We would now like to understand the conservation equations more covariantly and locally. This requires that we use a general vector field $\xi\in TB$, and combine the Raychaudhuri and Damour equations as components of one equation.

In order to decide which terms on the RHS of (\ref{Ray1}) are part of the boundary current and which are part of the fluxes, let us recall the form of the energy-momentum flux for a scalar field with Lagrangian $L=\frac12 g^{ab}\pa_a \phi\pa_b\phi - V(\phi)$. On a null surface, the canonical momentum density $P$ conjugate to $\phi$ is $P = L[\phi] \eb$, and the energy momentum tensor becomes
\be
T_{LL} \eb = L[\phi]L[\phi]  \eb, \qquad T_{Lv} \eb = L[\phi] v[\phi] \eb.
\ee
Those components combine into $T_{L\xi} \eb = L[\phi] \xi[\phi] \eb$ for $\xi \parallel B$. For a scalar field, the flux that controls the flow of energy and momenta through $B$ thus has a natural canonical expression given by the product of the momenta with the field transforms
\begin{align}
	T_{L \xi} \eb = P {\LL}_\xi\phi.
\end{align}
This canonical expression is generic to any form of matter.
We therefore expect the gravitational flux term to have a similar canonical form $\sum_i P_i \LL_\xi Q_i$.

In order to establish this we  need to isolate terms that can be interpreted in a  canonical form $P \LL_\xi Q$, from the equations (\ref{Ray1}). Lets first recall the action of diffeomorphisms on our data (section \ref{ssec:diffeos}): We have
\begin{align}
	\LL_\xi \gamma_{AB} ={}& 2\big( f \sigma_{AB} + e^{-2\varphi} \rd_{<A} v_{B>}\big), &\LL_\xi \eb ={}& (f\theta+\rd_A v^A) \eb,\\
	\LL_\xi L^a ={}& [v, L]^a, & \qquad \LL_\xi \mu ={}&  L[  (L+\mu)[f]] + v[\mu] .
\end{align}
We can now express the Raychaudhuri equation contracted with $f L^a$ as a canonical conservation law. Using again $\L_L (g \eb) = \rd (g \es) = (L[g] + \theta g)\eb$, we get
\begin{align}
	(fL)^a G_{a L} \eb ={}& - \rd (f \theta \es) + (f \mu + L[f]) \rd \es - f \sigma_a{}^b \sigma_b{}^a \eb.
\end{align}
The second term on the RHS is not of the canonical form $P \LL_\xi Q$, so we integrate by parts and use $(\rd g) \wedge \es = L[g] \eb$ to get
\begin{align}
	(fL)^a G_{a L} \eb ={}& \rd \big( - f \theta \es + (L+\mu)[f] \es \big) - L[(L+\mu)[f]] \eb - f \sigma_a{}^b \sigma_b{}^a \eb\\
	={}& \rd \big(  (L+\mu-\theta)[f] \es \big) -  \eb( \LL_{f L} \mu + \tfrac12  \sigma^{AB} \LL_{fL} \gamma_{AB}).
\end{align}
The RHS is now written as the sum of the  differential of a $(D-2)$-form and two canonical flux terms, this is the form we want.

Let us turn to the densitized Damour equation (\ref{Damour1}).
Contracting with $v^a$ and using $\rd_a v^a \eb = \rd ( \iota_v \eb)$, we can rewrite it as
\begin{align}
	v^a G_{La} \eb \overset{B}={}& \rd \big(v^a \omega_a \es+ (v^a \sigma_a{}^b) \iota_b\eb\big) + \big(\omega_a [v,L]^a  - (v^a \rd_a \mu) - (e^{-2\varphi}\sigma^{ab} \rd_a v_b)\big) \eb\\
	={}& \rd \big(v^a \omega_a \es+ (v^a \sigma_a{}^b) \iota_b\eb\big) - \eb  
	(\LL_v \mu -  \omega_a \LL_v L^a  - \tfrac12 \sigma^{AB} \LL_\xi \gamma_{AB}).
\end{align}
The RHS is also written as the sum of a differential plus  three canonical flux terms. 

We can now combine the Raychaudhuri and Damour equations and express the Einstein equations $G_{\xi L}= T_{\xi L}$ as a canonical conservation equation. Let us again parametrize $\xi = fL+v$, and define the boundary current  $j^a_\xi$ as
\bea\label{j}
\boxed{j_\xi^a :={} \big( (L+ \mu-\theta)[f] + v^b \omega_b\big) L^a +  v^b \sigma_b{}^a.}
\eea
The corresponding boundary current form  is the $(D-2)$-form $J_\xi := \iota_{j_\xi} \eb$ given by
\be\label{J} \boxed{
J_\xi  = (L + \mu - \theta)[f] \es + v^b \omega_b \es + v^b \sigma_b{}^a \iota_a \eb.}
\ee
Setting $G_{ab} = T_{ab}$, we can then write the Raychaudhuri and Damour equations as
\begin{equation}\label{conservation equation}
\boxed{\begin{aligned}
	\rd J_\xi \overset{B}={}& \big( T_{\xi L}+  \tfrac12 \sigma^{ab}({\LL}_{\xi}\gamma_{ab})
	- \omega_a ({\LL}_{\xi} L^a) +{\LL}_{\xi}\mu \big) \eb.
	\end{aligned}}
\end{equation}
In this expression the gravitational flux is now expressed in a canonical form.
The equations (\ref{J}, \ref{conservation equation}) summarize the null gravitational constraint equations. The expression for the boundary current $J_\xi$ is determined by this analysis up to a total differential $J_\xi \to J_\xi + \rd \beta_\xi$. 

The gravitational flux terms, which appear alongside the matter flux terms on the RHS of (\ref{conservation equation}), are of the canonical form $P \LL_\xi Q$. The canonical pairs are usually identified using the symplectic potential or related technology, but this analysis provides an alternative route towards their identification. We see that the gravitational canonical pairs $(P, Q)$ are the spin-2 pair $(\tfrac12\sigma^{AB} \eb, \gamma_{AB})$ of densitized shear and conformal metric, the spin-1 pair $(- \omega_a \eb, L^a)$ consisting of the twist and the null directions, and the spin-0 pair $(\eb, \mu)$ consisting of the area form  and  spin-0 momentum.

Let us interpret the boundary current vector (\ref{j}). In a given reference frame, the time component of a current vector is interpreted as the charge density and the spatial components as non-relativistic currents. In analogy, we may interpret the components along $L$ of $j_\xi^a$ as charge aspects. First, consider a vector field $\xi = fL$ parallel to $L$, which we interpret as a ``null time'' translation.  The conserved charge of time translations is energy, and we thus find the \emph{gravitational energy aspect}
\begin{align}\label{eq:e_aspect}
	\boxed{ e_f = (- \theta + \mu + L)[f] \es.}
\end{align}
It can be rewritten as $e_f = (\kappa - \tfrac{1}{D-2} \theta + L)[f] \es$, and features the combination $\kappa - \tfrac{1}{D-2} \theta$, which also appears in the non-densitized Raychaudhuri equation. 
Note that the gravitational energy aspect $e_f$ differs from the previous energy density $-f \theta\eb$ by the addition of a pressure term $\mu_f \eb$ with $\mu_f := f \mu + L[f]$. We can therefore interpret $-f \theta\eb$ as an internal energy of the sphere $S$ while $e_f$ is its enthalpy.


The conserved charge for spatial vector fields is the momentum. We can thus identify from (\ref{j}) the \emph{momentum aspect} 
\begin{align}
	\boxed{	p_v = v^b \omega_b \, \es.}
\end{align} 
The term $v^b \sigma_b{}^a$ then finds interpretation as a spatial momentum current. This current can be related to soft currents as they appear at null infinity \cite{Strominger:2017zoo}.

In (\ref{j}), we have written the boundary current $j_\xi$ using the split $\xi \rightarrow (f, v)$ and the extrinsic geometry of $B$. It can also be written more covariantly and geometrically if we recognize (see eq. \ref{eq:nabla_L_xi_into_B})  that 
\begin{align}
\nabla_L \xi^a = - [v, L]^a + \big((\kappa + L)[f] + v^b \omega_b\big)L^a + v^b \theta_b{}^a,
\end{align}
and use $\kappa - \tfrac1{D-2} \theta = \mu-\theta$. We get 
\bea\label{j2}
\boxed{
	j_\xi^a = \nabla_L \xi^a  - \tfrac1{D-2} \theta \xi^a + [v, L]^a.
}
\eea
The dependence on the extrinsic geometry of $B$ is now captured by the spacetime covariant derivative $\nabla_L \xi^a$.

To summarize, in (\ref{conservation equation}) we have rewritten the null Raychaudhuri and the Damour equations as a conservation law on the null surface $B$, equating the divergence of the gravity boundary current \ref{J} to the matter energy-momentum flux $T_{\xi L}$ plus a gravitational flux  of the canonical form $\sum_i P_i \LL_\xi Q_i$.

\section{Charges from Symplectic Potential}\label{sec:from_sp}
We will now give a more canonical derivation of the conservation equation, 
starting from the explicit expression for the null gravity symplectic potential in terms of the intrinsic and extrinsic geometry of $B$ derived in \cite{Hopfmuller:2016scf} and using technology from the covariant Hamiltonian formalism (see, e.g., \cite{Crnkovic:1986ex, Lee:1990nz, Donnelly:2016auv, Speranza:2017gxd}).

\subsection{Covariant Hamiltonian Formalism and its Ambiguities}
Given a total Lagrangian density $L^T = L + L^M$, where $L$ is the gravity Lagrangian and $L^M$ the matter Lagrangian, the \emph{symplectic potential current} $\Theta^T$ is defined implicitly as
\be\label{eq:def_Theta}
\delta L^T = - E^T + \rd \Theta^T
\ee 
It is a one-form on field space, i.e., has one $\delta$. Morally, when pulled back onto a hypersurface, the core piece of $\Theta^T$ is of the form $\sum P \delta Q$ and allows to read off the canonical configuration and momentum variables for the matter and gravity sector. Specific details  and boundary ambiguities, however, will be important. 
Here we work with the Einstein-Hilbert Lagrangian density $L =  \frac12 \epsilon (R - 2 \Lambda)$ with $\epsilon$ the spacetime volume element, and an arbitrary minimally coupled matter Lagrangian $L^M$. The equations of motion are $E^T = \frac12 \epsilon (G^{ab} + \Lambda g^{ab} -T^{ab}) \delta g_{ab}$, with $T^{ab}$ the gravitational matter energy-momentum tensor (we work in units where $8\pi G=1$).

For a diffeomorphism covariant Lagrangian density $L^T$, the \emph{Noether boundary current} $J^T_\xi$ on spacetime $M$ is defined by:
\be\label{eq:NT}
I_\xi \Theta^{T} - \iota_\xi L^T  \,{=}\, C_\xi + \rd J^T_\xi.
\ee
Here $I_\xi$ is the contraction on field space, i.e., $I_\xi \Theta^T(g_{ab}, \delta g_{ab}) = \hat \Theta^T(g_{ab}, \L_\xi g_{ab})$, while $\iota$ is the space-time contraction. The Noether boundary current $J^T_\xi$ is a $(D-2)$ form that can be integrated on spheres $S$. It is often   referred to as the gravitational ``superpotential''  \cite{Szabados2009}.

The LHS of (\ref{eq:NT}) is the Noether (bulk) current density associated with diffeomorphism symmetry. The first Noether theorem is the statement that this  current is conserved on-shell, i.e., $\rd(I_\xi \Theta^{T} - \iota_\xi L^T)=0$. The RHS of (\ref{eq:NT}) expresses this current as the sum of a bulk piece $C_\xi$ and a boundary piece $\rd J^T_\xi$. The bulk piece $C_\xi$ is the constraint $(D-1)$-form, and because diffeomorphisms are  gauge, it vanishes when the equations of motion are satisfied:
\begin{align}
	C_\xi = \xi^a (G_a{}^b + \Lambda \delta_a^b - T_a{}^b) \iota_b \epsilon \  \hat = {}\ 0.
\end{align}
This means that the Noether current is  a pure boundary term on-shell given by $\rd J^T_\xi$, and the integral over a $(D-2)$ sphere of $J^T_\xi$ gives the Noether charge contained within the sphere.

It is useful write the full Noether current density on the LHS of (\ref{eq:NT}) as the sum of a   matter contribution associated with $L^M$ and a gravitational contribution
associated with $L$. The matter contribution to the Noether current density is
\be
I_\xi \Theta^{M} - \iota_\xi L^M = T_{\xi}{}^a \epsilon_a - \rd J^M_\xi.
\ee
The LHS is, by definition, the canonical matter energy-momentum tensor, and the RHS contains the gravitational matter energy-momentum tensor $T_\xi{}^a$ and possibly a total derivative $\rd J^M_\xi$. $J^M_\xi$ is the matter contribution to the Noether boundary current.
The fact that there could be a difference between the canonical energy momentum tensor and the gravitational energy momentum tensor  is well known (see \cite{Blaschke:2016ohs} and references therein for an elementary review). This was recognized long ago by Belinfante \cite{Belifante1} and also enters the ``improvement'' needed in  order to describe properly conformal currents  \cite{Callan} that satisfy Ward identities. 

The presence of a  non-trivial edge mode contribution to the canonical energy-momentum tensor is due  to the presence of a spin current, which vanishes for scalar fields, but not for non-zero spin fields such as gauge fields. For Yang-Mills with Lagrangian $L^M =\tfrac1{g^2} \tr(*F\wedge F)$ the diffeomorphism boundary current $J^M_\xi$ coincides with the gauge Noether charge form associated with the gauge parameter $\iota_\xi A$ and reads $J^M_\xi = \tfrac1{g^2} \Tr(*F \iota_\xi A )$. It is thus natural to accompany  the infinitesimal diffeomorphism $\xi$ with a field dependent gauge transformation with parameter $- \iota_\xi A$. Under this combined transformation, the matter boundary current vanishes and the canonical and gravitational energy-momentum tensors agree. The total boundary current then only involves the gravity phase space variables. In the following, we assume that the matter boundary current has been taken care of and focus on the gravity charges.

When (\ref{eq:NT}) is pulled back on the null hypersurface $B$ and contracted with a vector $\xi$ tangential to $B$, the term $\iota_\xi L$ and the cosmological constant do not contribute, and if the canonical and gravitational matter energy-momentum agree, we get
\be\label{eq:NT_on_B}\boxed{
 \xi^a G_{aL} \eb = \jmath^*_B( \rd J_\xi - I_\xi \Theta),}
\ee
where $\Theta$ is the gravity symplectic potential. For the LHS, we have used $\jmath^*_B (\iota_b \epsilon) = - L_b \eb$.
Since the symplectic potential contains the terms $\Theta \overset{B}= P \delta Q$, we expect that $I_\xi \Theta = P \LL_\xi Q$ reproduces the flux terms of the last section. Then, we can identify the Noether boundary current $\jmath^*_B (J_\xi)$ with the boundary current of the conservation law \Ref{conservation equation}, and (\ref{eq:NT_on_B}) and (\ref{conservation equation}) become the same canonical conservation equation.

But importantly, the quantities we have used admit three ambiguities: the JKM ambiguities \cite{Jacobson:1993vj}.
Firstly, the Lagrangian can be shifted by a total differential $\rd \l$ which changes the symplectic potential by a total variation. This corresponds to a canonical transformation, or a change of polarization. Secondly, it is clear from its implicit definition (\ref{eq:def_Theta}) that the symplectic potential is defined only up to a closed $(D-1)$-form, it can hence be shifted by an exact form $\rd \alpha$ which is a one-form on field space, i.e., contains one $\delta$. These two ambiguities send
\be
L\to L+ \rd \ell,\qquad  \Theta \to \Theta + \rd \alpha +\delta \ell. 
\ee
And the  Noether boundary current feels these ambiguities: Assuming that the boundary action $\ell$ is covariant\footnote{i.e., $\LL_\xi \ell= {\cal L}_\xi \ell$, at least for $\xi \parallel B$. Note that the existing proposals for a Gibbons-Hawking like $\l$ \cite{Lehner:2016vdi, Parattu:2016trq} involve $\kappa \eb$ and are not covariant in this sense, so there is an extra term in (\ref{eq:modifying_Q}).}, using (\ref{eq:NT}) it becomes
\be\label{eq:modifying_Q}
J_\xi \to J_\xi + \iota_\xi \ell + I_\xi \alpha. 
\ee
As the third ambiguity, since $J_\xi$ is also defined implicitly through (\ref{eq:NT}), it may be shifted by a closed $(D-2)$-form that depends on $\xi$.

In the next subsection we fix the ambiguity $L \rightarrow L + \rd \l$ by working with the Einstein-Hilbert Lagrangian, see section \ref{sec:hams} for an argument favoring that choice. The ambiguity $\Theta \rightarrow \Theta + \rd \alpha$ will be fixed demanding that there are no corner symplectic pairs on $\p B$. As it will turn out, we can then express the RHS of (\ref{eq:NT_on_B}) entirely in terms  the intrinsic and extrinsic geometry of the null surface $B$.

\subsection{Null Symplectic Potential and Intrinsic Symplectic Potential}
Starting from the Einstein-Hilbert density $L$, via (\ref{eq:def_Theta}) we obtain the well-known standard symplectic potential current
\begin{align}\label{eq:theta_standard1}
	\hat\Theta[g, \delta g] = \tfrac12\nabla_b (\delta g^{ab} - g^{ab}  \delta g)   \epsilon_a ,
\end{align}
where $\epsilon = \sqrt{g} \rd^n x $ is the volume $D$-form, 
$\epsilon_a = \iota_{\p_a} \epsilon$ is the directed codimension 1 volume element and $\delta g^{ab}:= g^{aa'} g^{bb'} \delta g_{a'b'}$ is the variation of the metric, with trace $\delta g$. 

The standard symplectic potential current is covariant in the sense that it does not make reference to any background structure. This covariance property can be formalized by extending the definition of anomaly to field space forms. One considers  (for more details on this technology, see, e.g., \cite{Donnelly:2016auv, Speranza:2017gxd}):
\begin{align}\label{eq:def_anom_on_forms}
	\Delta_\xi  = \LL_\xi -\L_\xi - I_{\delta \xi}.
\end{align}
The term $I_{\delta \xi}$ compensates for possible dependence of the vector field $\xi$ on the metric, and the field space Lie derivative is extended to field space forms via the Cartan formula $\LL_\xi = \delta I_\xi + I_\xi \delta$ (remember that $\delta$ is an exterior derivative, so it involves antisymmetrization). Then $\hat\Theta$, as a $(D-1)$-form on spacetime $M$, has vanishing anomaly: $\Delta_\xi \hat\Theta = 0$ for all $\xi$.

The pullback of $\hat{\Theta}$ onto the null surface $B$ along the embedding $\jmath_B: B \hookrightarrow M$ can be rewritten in terms of the metric parameters (\ref{eq:gf_metric}) and the extrinsic geometry of $B$. This was done in \cite{Hopfmuller:2016scf}.
In the appendix \ref{app:symplectic_potential} we bring the result into the form which is useful here. The symplectic potential current $\hat\Theta$ on $B$ becomes the sum of a bulk and a corner term: 
\begin{align} \label{eq:standard_sp}
	\jmath_B^* \hat{\Theta} =
	\Theta - \rd \alpha_S
\end{align}
The bulk term is
\begin{align}\label{intriscpot}
\boxed{\Theta ={}  \left(\tfrac12 \delta \gamma_{ab}  \sigma^{ab} - \omega_a \delta L^a  + \delta \mu\right) \eb}
\end{align}
We recognize the canonical $P$s and $Q$s appearing in the flux terms of the conservation law (\ref{conservation equation}). The boundary contribution is
\begin{align}\label{eq:alpha_S}
\alpha_S  ={}&  \tfrac12 \left( \delta h \, \es + \imath_{\delta L} \eb  \right) 
- \tfrac1{D-2} \delta \es.
\end{align}
Recall that  $e^h$ is the scale of the  normal metric, $e^h  = \sqrt{|g|}/\sqrt{q}$ in the parametrization (\ref{eq:H_param}). The expressions (\ref{eq:standard_sp}) is valid for variations $\delta g_{ab}$ of the metric that keep the surface $B$ null, i.e. such that $\delta \beta \overset{B}{=} 0$ in the parametrization (\ref{eq:H_param}).

The boundary contribution can be rewritten as follows: using $\es =\iota_L \eb$ and $\delta \eb = (D-2) \delta \varphi \eb$ we have 
$
\delta \es  = \imath_{\delta L} \eb + (D-2) \delta \varphi  \es
$,
so 
\be
\alpha_S  = \frac12 \left(\delta h-  2 \delta \varphi\right) \, \es   
+ \tfrac12\tfrac{D-4}{D-2} \,  \imath_{\delta L} \eb.
\ee
It is interesting that the combination $h- 2 \varphi$ is invariant 
under local rescaling of the metric, and that the extra contribution vanishes in dimension $D=4$.

As we have seen the symplectic potential current is defined up to the addition of a closed form. That means that the corner term $\alpha_S$ can be removed by exploiting the ambiguity. We define the \emph{intrinsic symplectic potential current} on $B$ as \begin{equation}\boxed{
	\Theta := \hat\Theta + \rd \alpha_S.}
\end{equation}
It coincides with the bulk piece (\ref{intriscpot}).
This choice fixes the closed ambiguity in the symplectic potential in such a way that the modified symplectic potential has no boundary pairs. 
Our symplectic potential current contains only the intrinsic geometry $\{\gamma_{ab}, L^a, \varphi \}$ and extrinsic geometry $\{\sigma^{ab}, \omega_a, \mu \}$ of $B$. This is in contrast to the usual expression (\ref{eq:standard_sp}) that also contains $\delta h$, which fits in neither of these categories. We choose the name ``intrinsic'' because as we saw in section \ref{ssec:diffeos}, the transformation under diffeomorphisms of the data contained in the intrinsic symplectic potential does not depend on how the diffeomorphism is extended outside $B$. Again, $h$ is not intrinsic in this sense.

There exists already in the literature some discussion on the closed ambiguity, in the case where one choses particular boundary conditions and demands that the symplectic form should be conserved, i.e., independent of which Cauchy surface is chosen. For instance this  was done in \cite{Campiglia:2017mua} in the context of electromagnetism at spatial infinity,
and in \cite{Ashtekar:2000hw},    in  the context of gravity for isolated horizon boundary conditions\footnote{Incidentally, the corner modification used in \cite{Ashtekar:2000hw} is similar to ours.}. 

Let us take a closer look at the three terms  that appear in the boundary contribution (\ref{eq:alpha_S}). Removing the first term  $- \tfrac12 \rd (\delta h \es)$ from $\hat \Theta$ is central to our analysis, as we will see in the next subsection. The second term term $- \tfrac12\rd (\iota_{\delta L} \eb)$ does not enter the integral $\int_B \hat\Theta$ if the boundaries $\p B$ are aligned with the foliation $S$, since $\delta L$ is parallel to $S$. Its removal thus does not influence the boundary current integrated on $S$, but rather modifies the parts of the boundary current that vanish when pulled back to $S$\footnote{More specifically, it talks to the term $[v, L]$ in \ref{j2}.}. Lastly, the term $\rd \tfrac{1}{D-2}\delta \es$ is both a total derivative and a total variation, it could thus also be understood as arising from a codimension two corner action proportional to the corner area. Removing it does not change the symplectic form.

Having fixed the ambiguity of the symplectic potential current, we can now evaluate the flux term $I_\xi \Theta$ on the RHS of (\ref{eq:NT_on_B}). It becomes
\begin{align}\label{eq:canonical_flux_term}
	\boxed{I_\xi \Theta = \big(\tfrac12 \sigma^{AB} \LL_\xi \gamma_{AB} - \omega_a \LL_\xi L^a + \LL_\xi \mu\big)\eb,}
\end{align}
which coincides with the flux term from the conservation equation (\ref{conservation equation}).
From the intrinsic symplectic potential we can express the kinematical Poisson brackets as
\begin{align}
\{ e^{(D-2) \varphi}\sigma^{AB}(x), \gamma_{CD}(y) \} =& 2 \delta(x,y) \delta^A_C\delta^B_D,\cr
\{ U^A(x), \omega_B(y) e^{(D-2) \varphi}\}=&\delta(x,y) \delta^A_B\cr
\{ \varphi(x),\mu(y)\}=& \tfrac1{D-2}  \delta(x,y)
\end{align}
where $ \delta(x,y)$ is the Dirac delta distribution for the coordinate measure $\rd \phi^0 \wedge \rd^{D-2}\sigma$ on $B$.

\subsection{Noether Charge and Conservation Law}

Let us now turn to the boundary current term on the RHS of  (\ref{eq:NT_on_B}). To evaluate it, we start from the boundary current $\hat J_\xi$ of the standard symplectic potential (\ref{eq:theta_standard1}). $\hat J_\xi$ is the well-known Komar charge form \cite{Komar:1958wp}, which is a $(D-2)$-form on spacetime $M$, and given by
\be
\hat J_\xi=  \tfrac12 *\rd g(\xi) = \tfrac12 \epsilon_{ab} \nabla^a \xi^b,
\ee 
where $\epsilon_{ab} = \iota_a \iota_b \epsilon$.
Pulling back onto $S$ and parameterizing $\xi = fL + \bar f \Lb + v$, one gets
\begin{align} \label{eq:noether_charge_std}
	i_S^* (\hat J_\xi) ={}& \tfrac12 ( (L + \kappa) [f] - (\Lb + \bar\kappa) [\bar f] + (\omega_a - \eta_a) v^a)\rd S,
\end{align}
where as before $\eta_a = - q_a{}^b \nabla_\Lb L_b$ and $\bar\kappa = L_a \nabla_{\Lb} \Lb^a$.
The Komar charge form has the advantage of being simple, and covariant under all diffeomorphisms. It is the charge most commonly used in canonical analyses, see, e.g., \cite{Iyer:1994ys}. However, it possesses two features that make it unsatisfactory for the analysis of conservation laws along a null hypersurface $B$: 
It depends not only on $\xi$ as a vector field on $B$, but on its extension outside of $B$ through the transverse derivative $\Lb[\bar f]$. Therefore even a vector field which vanishes on $B$ may have non-zero charge.
We also see that in addition to the variables $\kappa, \omega_a$ which form part of the extrinsic geometry of $B$ as embedded in $M$,  $\hat J$ involves the variables $\bar\kappa$ and $\eta$ which cannot be interpreted in terms of the intrinsic or extrinsic geometry of $B$. In trying to describe physics from the viewpoint of the null surface $B$, both these features are undesirable.

We now show directly that the Noether boundary $J_\xi$
associated with  the intrinsic symplectic potential $\Theta_B$ (\ref{eq:canonical_flux_term}) resolves both issues affecting the Komar boundary current: The edge mode current $J_\xi$ does not contain derivatives of $\xi$ transverse to $B$, so it vanishes if $\xi$ vanishes on $B$. Furthermore, $J_\xi$ is entirely determined by the intrinsic and extrinsic geometry of $B$.
We also show that $J_\xi$ coincides, up to a total differential, with the boundary current (\ref{J}) which we found from analyzing the constraints. 

By (\ref{eq:modifying_Q}), the boundary current $J_\xi$ of the modified symplectic potential is related to the Komar charge form $\hat J_\xi$ as
\be\label{Improved}
J_\xi = \hat J_\xi + I_\xi \alpha_S.
\ee
The core reason ensuring the properties of $J_\xi$ is that we removed $\delta h$ from the symplectic potential current. Using the transformation of $h$ given in (\ref{eq:LL_xi_h}), we have
\begin{align}
	I_\xi \big(\tfrac12 \delta h \rd S\big) = \tfrac12 \big( (L + \kappa) [f] + (\Lb + \bar\kappa) [\bar f] + (\omega_a + \eta_a) v^a \big) \rd S.
\end{align}
It is then clear that adding this term to (\ref{eq:noether_charge_std}) removes both the transverse derivative acting on $\bar{f}$  as well as the dependence on $\bar\kappa$ and $\eta_a$.

In detail, $J_\xi$ is obtained as follows: As a form on $B$, the Komar charge form reads
\begin{align}\label{eq:Komar_on_B}
	\jmath^*_B (\hat J_\xi) ={}& \tfrac12 L_a (\nabla^a \xi^b - \nabla^b \xi^a) \iota_b \eb,
\end{align}
where we used $\jmath^*_B (\epsilon_{ab})= L_a (\iota_b \eb) - L_b (\iota_a \eb)$. The expression $I_\xi \alpha_S$ is computed most efficiently by rewriting $\alpha_S$ as
\begin{align}\label{alphaf}
\alpha_S = \left(\frac12 L^{a} \delta g_{ac} g^{bc}  + {\delta L}^b \right)  \iota_{b} \eb  - \tfrac{1}{D-2} \delta (\es).
\end{align}
To pass from (\ref{eq:alpha_S}) to the last line, use that $L^a \delta g_{ac} g^{bc} = \delta (L_c) g^{bc} - \delta (L^b)$. Consulting (\ref{eq:dualpair}) one sees that on $B$, $\delta (L_c) = \delta h L_c$. To compute $I_\xi \alpha_S$, further recall from \ref{ssec:diffeos} that for $\xi = fL + v$ we have 
\begin{align}
	I_\xi \delta \es ={}& \L_\xi \es = \iota_\xi (\theta \eb) + \rd (\iota_\xi \es)\nonumber \\
	I_\xi \delta L^a ={}& [v, L]^a.
\end{align}
Using also $I_\xi \delta g_{ab} = \nabla_a \xi_b + \nabla_b \xi_a$, altogether we get
\begin{align}\label{eq:I_xi_alpha}
	I_\xi \alpha_S ={}& \frac12 L_a (\nabla^a \xi^b + \nabla^b \xi^a) \iota_b \eb + [v, L]^a \iota_a \eb - \tfrac{1}{D-2} (\iota_\xi \theta \eb + \rd \iota_\xi \es).
\end{align}
Adding (\ref{eq:Komar_on_B}) and (\ref{eq:I_xi_alpha}) yields the Noether boundary current $J_\xi = \hat J_{\xi} + I_\xi \alpha_S$:
\begin{align}\boxed{
	J_\xi = \big( (\nabla_L \xi^a) + [v, L]^a \big) \iota_a \eb - \tfrac{1}{D-2} \iota_\xi \theta \eb - \rd (\tfrac{1}{D-2} \iota_\xi \es)}
\end{align}
As claimed, this coincides with the boundary current (\ref{J}) that we found analyzing the constraints, up to the total derivative term $\rd (\tfrac{1}{D-2} \iota_\xi \es)$, which is within the ambiguity of both prescriptions and vanishes when integrated on any closed surface.

Let us reiterate the results of this section: The general identity (\ref{eq:NT_on_B}) on $B$ reads
\begin{align}
 G_{\xi L} \eb ={}& \rd J_\xi - I_\xi \Theta.
\end{align}
This equation coincides exactly with the conservation equation (\ref{conservation equation}) for the edge mode current --- if we use the intrinsic symplectic potential current $\Theta$ of (\ref{intriscpot}), which differs from the standard one by a total derivative and has no corner pairs. The gravitational flux terms are given by $I_\xi \Theta$ and are of the same form as the canonical energy-momentum of matter, i.e., $P \LL_\xi Q$. The boundary current $J_\xi$ is (essentially) given by the Noether boundary current of the intrinsic symplectic potential current $\Theta$. As a further consequence of modifying the symplectic potential current, everything is expressed in terms of the intrinsic and extrinsic geometry of the null surface $B$, and independent of the extension of the vector field $\xi$ outside of $B$.

\section{Hamiltonians}\label{sec:hams}

What makes conserved charges interesting, especially in the quantum theory, is that they usually are the generators, i.e., the Hamiltonians, of symmetries. We now turn to the question how the charge $\int_{\p B} J_\xi$ is related to the Hamiltonian generating the infinitesimal transformation $\xi$. See \cite{Kijowski1997} for a related discussion.

The Hamiltonian for $\xi$, if it exists, should satisfy
\begin{align}
	\{H_\xi, F\} = \LL_\xi F,
\end{align}
for any functional $F$ of the metric. The Poisson brackets are given in terms of the inverse symplectic form (which is a bivector on field space, and of course ill-defined before performing the symplectic reduction) as $\{F, G\} = \Omega^{-1} (\delta F, \delta Q)$. Using that $F$ is arbitrary and contracting both sides with the symplectic form $\Omega := \int_B \delta \Theta$ one obtains the equivalent expression
\begin{align}\label{eq:deltaH}
	\delta H_\xi = - I_\xi \Omega,
\end{align}
where $\Omega = \int_B \delta \Theta$ is the symplectic form, and again $I$ is the contraction on field space. In general, there is no guarantee that $- I_\xi \Omega$ is an exact variation and thus that a Hamiltonian exists.

In order to see the relation between Hamiltonians and charges, let us calculate the RHS of (\ref{eq:deltaH}), focusing just on the gravity sector. We are interested in the on-shell Hamiltonian, and dropping all terms involving the constraints  yields (see appendix \ref{ap:ham})
\begin{align}\label{eq:ham_for_anomalous}
	-I_\xi \Omega \hat = \int_{ \p B} \Big(\delta (J_\xi) - J_{\delta \xi} - \iota_\xi \Theta - \Delta_\xi \alpha_S \Big).
\end{align}
To proceed further, let us fix the metric dependence of  $\xi$, and set $\xi = f L + v$ with $\delta f = 0$ and $\delta v = 0$, such that $\delta \xi = f \delta L$. Intuitively this means that the direction of $\xi$ relative to the null generators is fixed, or from a fluid perspective that $\xi$ is fixed in a ``co-moving'' frame. As before, we also assume that the boundaries $\p B$ are aligned with the cross-sections $S$. Using that $i_S^*(\iota_\xi \eb) = f \rd S$, the ingredients are\footnote{Recall that $\rd S$ is the induced volume element on a cross-section $S$ of $B$.}:
\begin{align}
	i_S^* \delta (J_\xi) ={}& \delta \Big( \rd S \big( (- \theta + \mu)f + L[f] + v^a \omega_a\big)\Big)\nnn
	- i_S^* (J_{\delta \xi}) ={}& - f \delta L^a \omega_a \rd S \nnn
	- i_S^* (\iota_\xi \Theta) ={}& - f \left(\frac12 \sigma^{ab} \delta \gamma_{ab} - \delta L^a \omega_a + \delta \mu\right)\rd S \nnn
	- i_S^* (\Delta_\xi \alpha_S) ={}& - \delta L[f] \rd S.
\end{align}
For the last line, see appendix \ref{ap:ham}. Let us denote $\mu_f = f \mu + L[f]$. $\mu_f$ can be understood as the spin-0 momentum in the frame of the observer $\xi$. Noting $\delta \mu_f = f \delta \mu + \delta L[f]$ we get the following, remarkably compact result:
\begin{align}\label{eq:I_xi_Omega}\boxed{
	- I_\xi\Omega = \delta \left(\int_{\p B}  J_\xi \right) - \int_{\p B} \left(\tfrac12 f  \sigma^{ab} \delta \gamma_{ab} + \delta \mu_f\right)\es .}
\end{align}
The transformation  $\LL_\xi$ is a Hamiltonian symmetry with Hamiltonian $H_\xi$ if and only if the RHS is a total variation $\delta H_\xi$. Due to the presence of the second term we see that  boundary conditions are needed to ensure the existence of a Hamiltonian. The simple form of (\ref{eq:I_xi_Omega}) is a consequence of using the modified symplectic structure which has the corner pair $(h, \es)$ removed. A similar equation has also been given from a first order perspective in \cite{Wieland:2017zkf}.

Let us analyze the result (\ref{eq:I_xi_Omega}) for some different cases. First consider a ``superrotation-like'' transformation, i.e., a vector field $v$ which is parallel to the foliation $S$ to first order around $\p B$. Then, no boundary conditions are needed, and the Hamiltonian is just the charge:
\begin{align}
H_v = \int_{\p B} J_v = \int_{\p B} v^a \omega_a \es.
\end{align}
The charge, the momentum conjugate to the null directions $L$ and the Hamiltonian coincide. The situation is analogous to electromagnetism, where on a null surface, $F_{ru}$ plays the triple role of the momentum conjugate to $A_u$, the conserved Noether charge and the Hamiltonian generating gauge transformations. In our case, this simple situation is a consequence of using the modified symplectic potential without boundary pairs.

Next, consider null dilatations that ``stretch'' the null surface in the null direction at its corners, i.e., $\xi = f L$ with $f = 0$ at $\p B$. We get $\delta \mu_f = (\delta L)[f] = 0$, since $\delta L$ is parallel to $\p B$, so
\begin{align}
 H_{fL} ={}& \int_{\p B} L[f] \es.
\end{align}
The null dilatations are thus generated by the corner area element, 
they are Hamiltonian symmetries even if no boundary conditions are fixed.

The case of null translations $\xi = f L$ with $f$ non-vanishing at the corners $\p B$ is more subtle. Boundary conditions are needed to ensure the existence of a Hamiltonian. The boundary conditions can be split up into conditions on the pair $(\sigma^{ab}, \gamma_{ab})$ and the pair $(\epsilon_S, \mu_f)$. No boundary conditions are needed for the spin-1 pair $(L^a, \omega_a)$, this is because we have chosen $\xi$ to vary with $L$.

For the spin-2 pair $(\sigma^{ab}, \gamma_{ab})$, a possible boundary condition is fixing the shear $\sigma^{ab} = 0$ at $\p B$. That is done, e.g., at isolated, Killing and conformal Killing horizons, and in the far past and far future of future null infinity.
More generally we can impose at the boundary of $B$ any relationship of the form $\sigma_{ab} = F(\gamma_{ab})$. Alternatively, one can fix the conformal metric to be the conformal metric of the unit sphere such that $\delta \gamma_{ab} = 0$. Note that in four spacetime dimensions, $\p B$ is two-dimensional. If it has spherical topology, every metric is diffeomorphic to a metric conformal to the unit sphere metric. Thus, fixing $\gamma_{ab}$ can be interpreted as a condition on the coordinates, rather than on the metric degrees of freedom. The residual transformations preserving the condition are the conformal Killing vectors of the unit sphere. 

For the spin-0 pair $\epsilon_S \delta \mu_f$, one possible boundary condition is fixing the area element such that the term becomes the total variation $\delta (\epsilon_S \mu_f)$. This leads to a Hamiltonian for null translations
\begin{align}
H_\xi^{\text{area}} = - \int_{\p B} \theta f \es.
\end{align}
Under boundary conditions fixing the area element of the corner, the generator of translations along $L$ is minus the expansion\footnote{Note, however, that a sensible $\xi$ should preserve the frame conditions, so for $f L$ to preserve a fixed value of $\es$ we have to set $\theta = 0$.}. 
See \cite{Czuchry:2004dc} for related results. Fixing the area element can also be viewed as a condition on the location of the spheres $\p B$, rather than a condition on the metric, such as in Bondi gauge at null infinity. A null translation then has to be accompanied by a radial diffeomorphism to restore the size of the spheres.

As a more general spin-0 boundary condition, one could provide a constitutive relation linking $\epsilon_S$ and $\mu_f$. This situation arises in black hole thermodynamics \cite{Wald:1993nt}, where (\ref{eq:I_xi_Omega}) becomes the ``Hamiltonian first law'' of black hole thermodynamics.

As another spin-0 boundary condition, one can fix $\mu_f$. We remind the reader that $\mu = \kappa + \tfrac{D-3}{D-2}\theta$ and $\mu_f = f \mu + L[f]$. For an isolated horizon, where the expansion vanishes, this conditions amounts to fixing the horizon ``temperature'' $\kappa$.  
Since any value for $\mu_f$ can be reached by choice of the coordinate field $\phi^0$ or by choosing the coefficient $f$, fixing $\mu_f$ can be interpreted as a condition on the clock $\phi^0$ or the vector field $\xi$ rather than on the metric degrees of freedom. The most obvious choice is fixing $\mu_f = c$ with a fixed constant $c$. The residual transformations preserving this condition  satisfy $L[f]/f = - c$.  The null translation Hamiltonian for fixed $\mu_f$ becomes 
\begin{align}
 H_\xi = \int_{\p B} (\mu_f - \theta f) \epsilon_S.
\end{align}
It coincides with the energy aspect (\ref{eq:e_aspect}) which we found by analyzing the constraints.

Since fixing $\mu_f$ gives a condition on how the coordinates are extended around the corners $\p B$, while fixing the area element $\epsilon_S$ requires moving the corners, from the viewpoint of a null surface at finite distance fixing $\mu_f$ seems a more natural condition than fixing $\epsilon_S$. 
Now, it is well known that boundary conditions are linked to a choice of boundary action: the symplectic potential of the full action should be made to vanish by the boundary conditions. As we saw, the symplectic potential of the Einstein-Hilbert action contains $\delta \mu \eb$, which vanishes when $\mu$ is fixed. If a Gibbons-Hawking like null boundary action containing $\int_B \mu \eb$ is added, the term in the symplectic potential becomes $- (\delta \eb) \mu$, which vanishes if $\rd S$ is fixed. From the perspective of a null hypersurface at finite distance, it thus seems more natural to work with the pure Einstein-Hilbert Lagrangian, rather than adding a ``null Gibbons-Hawking'' boundary action to switch to the metric polarization.

Different conditions on the spin-0 sector have appeared in the literature. The ``time'' $\phi^0$ can be linked to the total area of the cross-section $S$ at $\phi^0$ as in \cite{Reisenberger:2012zq}, which fixes the expansion $\theta$. One can use an affine parameter along the null geodesics, which fixes $\kappa = 0$.  As stated earlier, the combination $\kappa - \tfrac{1}{D-2} \theta=\mu-\theta$ can be set zero to simplify the Raychaudhuri equation \cite{0264-9381-10-4-012}. For a generic expanding null surface, the condition $\mu_f = 0$ is different from all of those.

To summarize, we asked for which symmetries $\xi$ and under which boundary conditions there exists a Hamiltonian generating the symmetry, using the intrinsic symplectic form. For spatial transformations $\xi^a = v^a$, a Hamiltonian always exists and is given by the twist field $\omega_a$. Null dilatations are generated by the area element $\epsilon_S$. For null translations, spin-2 and spin-0 boundary conditions are needed for the existence of a Hamiltonian. The most natural spin-0 boundary condition seems to be fixing the spin-0 momentum $\mu$, and the resulting Hamiltonian is the energy aspect (\ref{eq:e_aspect}).

\section{Conclusion}

We have presented an analysis of the evolution of gravitational canonical charges along null surfaces. In particular, we have clarified the relationship between the Raychaudhuri and Damour equations and the symplectic analysis. We have fixed the boundary ambiguity of the symplectic potential in a way that allows a formulation intrinsic to the null boundary.
This work can be viewed as a necessary step in understanding more deeply the connection between soft modes and edge modes. We expect to come back to this issue in the near future and study the asymptotic limit of our construction. We also hope that our work may aid a more gauge invariant, geometrical intuition of the Hamiltonians and conservation laws at null infinity.

Our analysis reveals the importance of  a new element in the canonical analysis: the spin zero momentum $\mu$.
We have emphasized its interpretation as a boundary pressure term. It would be interesting to develop further the thermodynamical interpretation of this element, and of the different null Hamiltonians. In particular we have seen that when the shear $\sigma^{ab}$ and $\mu$ vanish, such that the gravitational energy flux vanishes, there is a Hamiltonian interpretation of the gravitational energy aspect. When this is not the case it is tempting to interpret the gravitational flux as given in (\ref{eq:cst_em_tensor}) as an entropy production term. That interpretation is possible only if the term is positive such that the second law is satisfied. This is automatic for the spin-2 contribution $\sigma_a{}^b\sigma_b{}^a$, and one sees that null surfaces have a unit shear viscosity in gravitational units. The spin-0 term $- \mu \theta$ on the other hand is not generically positive. It becomes positive once we introduce a bulk viscosity coefficient $\nu>0 $ by the equation $\mu=-\nu \theta$. This defines a new notion of viscous null surfaces  which are  thermodynamically stable and which we would like to call ``thermodynamical horizons''. We plan to study the physical properties and the relevance of such thermodynamical horizons in the near future.

Finally, we considered only the constraint components $G_{La}$ for $a$ tangential to $B$. A full analysis will require also the canonical understanding of the transverse constraints $G_{L\bar{L}}$. We expect this constraint to find interpretation as a relativistic generalization of the Young-Laplace equation \cite{Freidel:2014qya} for viscous bubbles, but a detailed analysis is necessary.

{\bf Acknowledgments:} The authors thank William Donnelly, Simone Speziale, Hal Haggard, Aldo Riello and Tommaso de Lorenzo for fruitful discussions.
Research at Perimeter
Institute is supported by the Government of Canada through the Department of Innovation, Science and Economic Development Canada and by the Province of Ontario through
the Ministry of Research, Innovation and Science.

\appendix

\section{Symplectic Potential}
\label{app:symplectic_potential}
This appendix derives the form of the symplectic potential used in the main text. We starting from an expression given in \cite{Hopfmuller:2016scf}.
From the equation (5.27) in \cite{Hopfmuller:2016scf} we read that 
$\jmath_B^*\hat{\Theta} = \Theta_B + \rd \theta_S$ where 
\be
{\Theta}_B= \left(\tfrac12 \delta q_{ab} \theta^{ab} -\bar{\eta}_a \delta L^a
+ \delta(\kappa+\theta) \right) \eb
\ee
is the bulk symplectic potential and the boundary potential is given by 
\be
\theta_S = \frac12 \left[ \left(\tfrac12 h L^a \delta q + (h-1) \delta L^a\right) \imath_a \eb
- \delta\left([h L^a]  \imath_a \eb \right) \right].
\ee
Let us perform a trace-traceless split in $\Theta_B$, using
\be
\theta^{ab}=e^{-2\varphi}(\sigma^{ab} + \tfrac1{D-2} \gamma^{ab} \theta) ,
\qquad  \delta q_{ab} = \delta( e^{2\varphi} \gamma_{ab}) = e^{2\varphi}( 2\delta \varphi \gamma_{ab} + \delta\gamma_{ab}),
\ee
therefore 
\bea
(\tfrac12  \delta q_{ab} \theta^{ab})\eb  &=& (\tfrac12  \delta \gamma_{ab} \sigma^{ab})\eb + 
\theta \delta \varphi  \eb \cr
&=& (\tfrac12  \delta \gamma_{ab} \sigma^{ab})\eb  + \tfrac1{(D-2)} \delta (\eb) \theta \cr
&=& (\tfrac12  \delta \gamma_{ab} \sigma^{ab})\eb  - \tfrac1{(D-2)} ( \delta \theta)\eb +  \tfrac1{(D-2)} \delta (\eb \theta ).
\eea
Plugging in:
\be\label{TB}
{\Theta}_B= \left(\tfrac12 \delta \gamma_{ab} \sigma^{ab} -\bar{\eta}_a \delta L^a
+ \delta(\kappa+ \tfrac{D-3}{D-2} \theta) \right) \eb +  \tfrac1{(D-2)} \delta (\eb \theta ) , 
\ee

The boundary term can be rewritten, using that $\tfrac12 \delta q\, \imath_a \eb = \delta (\imath_a \eb)$ as 
\bea
\theta_S &=&
\frac12  \left( (h-1) \delta L^a - \delta[h L^a] \right) \imath_a \eb
\cr
&=& - \frac12  \left( \delta h  L^a + \delta L^a   \right) \imath_a \eb\cr
&=& -\frac12 \left( \delta h  \imath_L \eb + \imath_{\delta L} \eb  \right)
\eea

Noting also $\rd \es = \theta \eb$ and using the definition (\ref{eq:def_mu}) of $\mu$, we get the equations (\ref{eq:standard_sp}), (\ref{eq:standard_bdy_sp}) used in the main text:
\begin{align}
 \jmath_B^* \hat \Theta ={}& \eb \left(\frac12 \sigma^{AB} \delta \gamma_{AB} - \delta L^a \omega_a + \delta \mu \right) -\frac12 \rd (\delta h \iota_L \eb + \iota_{\delta L} \eb).
\end{align}

\section{Derivation of Raychaudhuri and Damour equations}\label{ap:eeq}
This appendix derives the densitized Damour equation
\begin{align}
	q_a{}^b G_{Lb} \eb = q_a{}^b \L_L (\omega_b \eb) +(\rd_b \sigma_a{}^b- \rd_a \mu)  \eb
\end{align}
and the densitized null Raychaudhuri equation
\begin{align}
	G_{LL} \eb = - \L_L (\theta \eb)+ (\mu \theta - \sigma^a{}_b \sigma^b{}_a) \eb.
\end{align}
We define
\begin{align}
\eta_a :={}& - q_a{}^b \Lb^c \nabla_c L_b\\
a_a :={}& q_{ab} \nabla_L L^b\\
\bar\theta_{ab} :={}& q_a{}^{a'} q_b{}^{b'} \nabla_{a'} \Lb_{b'}
\end{align}
and recall $\bar\eta_a := - q_a{}^b L^c \nabla_c \Lb_b$, while $\omega_a = q_a{}^b \Lb_c \nabla_b L^c$ and $\mu = \kappa + \tfrac{D-3}{D-2} \theta$.
The tangential acceleration $a_a$ vanishes on $B$ since, $L$ is geodesic on $B$.

For the Damour equation, we have 
\begin{equation}
q_a{}^b G_{Lb} = q_a{}^b R_{Lb} = q_a{}^c (\nabla_b \nabla_c L^b - \nabla_c \nabla_b L^b).
\end{equation} 
For the first term, use the decomposition of the identity $\delta^a_b = q_a{}^b + L_a \Lb^b + \Lb_a L^b$ to obtain
\begin{align}
	\nabla_a L^b ={}& \Lb_a L^b \kappa  - L_a \eta^b + \omega_a L^b + \theta_a{}^b + \Lb_a a^b.
\end{align}
Taking an additional derivative and projecting with $q$ yields
\bea
q_a{}^{a'} (\nabla_b \nabla_{a'} L^b) &=& q_a{}^{a'} \Big((\nabla_L\Lb_{a'})  \kappa  - (\nabla_b L_{a'}) \eta^b + \omega_{a'} \nabla_b L^b + 
\nabla_L \omega_{a'} + \nabla_b \theta_{a'}{}^b + \nabla_b \Lb_{a'} a^b \Big) \cr
&=&- \bar{\eta}_a  \kappa  - \eta^b \theta_{ba} + \omega_a (\kappa+\theta) + 
q_a{}^{a'}\nabla_L \omega_{a'} + q_a{}^{a'}\nabla_b \theta_{a'}{}^b + \bar\theta_{ba} a^b \cr
&=&(\omega- \bar{\eta})_a  \kappa   + 
q_a{}^{a'}\nabla_L \omega_{a'} + \theta \omega_a + \rd_b \theta_{a}{}^b + \theta_{a}{}^b \bar\eta_b  + \bar\theta_{a}{}^b a_b,
\eea
where for the second line we used $\nabla_aL^a = \kappa + \theta$ and for the third line we used 
$\nabla_b \theta_a{}^b = \rd_b \theta_a{}^b + (\eta+\bar\eta)_b \theta_a{}^b$.
The latter follows from
\bea
\nabla_b v^b &=& q^{ab} \nabla_a v_b + L^a \Lb^b \nabla_a v_b + L^b \Lb^a \nabla_a v_b\cr
&=& \rd_a v^a -  \nabla_L\Lb^b  v_b -  \nabla_{\Lb} L^b  v_b
\eea
for $v^a = q^a{}_b v^b$.
Using that $a_a = 0$ and $\bar\eta_a = \omega_a$ on $B$ (shown in section \ref{ssec:extrinsic}, and using that $\theta_{a}{}^b =  (\sigma_{a}{}^b + \tfrac1{D-2} \gamma_{a}{}^b \theta)$, this gives
\begin{align}
	q_a{}^c \nabla_b \nabla_c L^b \overset{B}{=}{}& q_a{}^c \nabla_L \omega_c + \theta_a{}^c \omega_c + \theta \omega_a + \rd_b \sigma_a{}^b + \tfrac{1}{D-2} \rd_a \theta\\
	={}& q_a{}^b (\L_L + \theta) \omega_b + \rd_b \sigma_a{}^b + \tfrac{1}{D-2} \rd_a \theta.
\end{align}
For the second line, we used $q_a{}^b \L_L \omega_b = q_a{}^b \nabla_L \omega_b + \theta_a{}^c \omega_c$.
The other term in $q_a{}^b R_{Lb}$ is simply $- q_a{}^c \nabla_c \nabla_b L^b = - \rd_a (\theta + \kappa)$, yielding
\begin{align}
	q_a{}^b R_{Lb} ={}& q_a{}^b(\L_L + \theta) \bar\eta_b + \rd_b \sigma_a{}^b - \rd_a(\kappa + \tfrac{D-3}{D-2} \theta).
\end{align}
Finally, using $\L_L \eb = \theta \eb$, this may be written in the form
\begin{align}
	q_a{}^b R_{Lb} \eb ={}& q_a{}^b \L_L (\bar\eta_a \eb) + \rd_b \sigma_a{}^b \eb - \rd_a \mu \eb.
\end{align}

Let us turn to the null Raychaudhuri equation. We have
\begin{equation}
G_{L L} =  R_{LL} = L^a (\nabla_b \nabla_a L^b - \nabla_a \nabla_b L^b).
\end{equation} 
For the first term, we use that
\begin{align}
	L^a \nabla_b \nabla_a L^b ={}&  \nabla_b (\nabla_L  L^b) - (\nabla_b L^a) (\nabla_a L^b)\cr
	={}& \nabla_b (L^b \kappa - a^b) - 
	(  L^a (\kappa  \Lb_b + \omega_b)  + \theta_b{}^a + \Lb_b a^a )(  L^b (\kappa  \Lb_a + \omega_a)  + \theta_a{}^b + \Lb_a a^b )\cr
	={}& L[ \kappa] + \kappa \nabla_a L^a -\nabla_a a^a 
	-[ \kappa^2  + 2 \omega_b a^b + \theta_b{}^a\theta_a{}^b]\cr
	={}& L[\kappa] + \kappa \theta - \theta_b{}^a\theta_a{}^b -\rd_a a^a  - (\eta+\bar\eta + 2 \omega)_a a^a.
\end{align}
We used
\begin{align}
	\nabla_a L^b ={}&  L^b (\Lb_a \kappa + \omega_a)   + \theta_a{}^b 
	+ \Lb_a a^b -  L_a\eta^b,
\end{align}
and $\nabla_a a^a = \rd_a a^a + (\eta_a + \bar\eta_a) a^a$.
The second term in $G_{LL}$ is simply $L[\kappa + \theta]$ and taking the difference gives
\be
G_{L L} =-L[ \theta] + \kappa \theta - \theta_b{}^a\theta_a{}^b -\rd_a a^a  - (\eta+\bar\eta + 2 \omega)_a a^a.
\ee
Using that $a = 0$ on $B$ and splitting $\theta^a{}_b$ into trace and traceless part as before we get
\begin{align}
	G_{LL} \overset{B}{=}{}&  -L [\theta] + \kappa \theta - \sigma_b{}^a\sigma_b{}^a -
	\tfrac1{(D-2)}\theta^2 \\
	={}&  -(L  + \theta)[\theta]   + \mu \theta - \sigma_b{}^a\sigma_b{}^a.
\end{align}
Using $\L_L \eb = \theta \eb$, we get
\begin{align}
	G_{LL} \eb = -\L_L (\theta \eb) + \mu \theta\eb - \sigma_b{}^a \sigma_a{}^b \eb.
\end{align}
The form of this equation is sensitive to the normalization of the null normal $L$. Introducing $\mu_f = f \mu + L[f]$ for an arbitrary function $f$ on $B$, we can write the densitized Raychaudhuri equation for an arbitrary normalization of $L$:
\be
f L^a G_a{}^b L_b \eb = - \L_{fL}( \theta \eb)  + (\mu_f \theta  - f\sigma_b{}^a\sigma_b{}^a) \eb.
\ee

\section{Diffeomorphism Actions}\label{ap:diffeos}
This appendix derives the field space Lie derivatives of intrinsic and extrinsic geometry along a spacetime vector field $\xi$. Since we are interested in which pieces of geometry ``talk to'' the extension of $\xi$ outside of $B$, we fix $\xi\parallel B$ at $B$ but allow for an arbitrary extension outside of $B$, setting $\xi^a = f L^a + \bar f \Lb^a + v^a$ with $\bar f = 0$ on $B$.

\textbf{Intrinsic Geometry:}
\begin{itemize}
	\item $\LL_\xi L^a$: We have $-q^{ab} L^c \delta g_{bc} = - q^{ab} (\delta L_b - g_{ab} \delta L^b) = \delta L^a$. We used that the $\phi^0$-component of $L^a$ is fixed: $\delta L^0 = 0$, such that $q^a{}_b \delta L^b = \delta L^b$. We also used that $\delta L_a=\delta h L_a$ is normal to $S$. We get
	\begin{align}
		\LL_\xi L^a ={}& I_\xi \delta L^a =- q^{ab} L^c (\nabla_b \xi_c + \nabla_c \xi_b)\cr
		={}& - q^{ab} L^c (\nabla_b v_c + \nabla_c v_b)\cr
		={}& q^a{}_b (v^c \nabla_b L_c - L^c \nabla_c v^b)\cr
		={}& q^a{}_b [v, L]^b = [v, L]^a.
	\end{align}
	We used that $\nabla_a L_b$ is symmetric when pulled back to $S$, and that $[v, L]^a$ is tangential to $S$ since $L$ preserves the foliation $S$. Noting $\L_\xi L^a = -L(f) L^a + [v, L]$, we obtain the anomaly $\Delta_\xi L^a = - [fL, L]^a = L(f) L^a$.
	\item $\LL_\xi h$: We have $\delta h = \delta g_{ab} L^a \Lb^b$, as may be checked explicitly in coordinates. Then
	\begin{align}
		\LL_\xi h ={}& (L^a \Lb^b+ \Lb^a L^b) \nabla_a (f L_b + \bar f \Lb_b + v_b)\cr
		={}& (L + \kappa)[ f] + (\Lb + \bar\kappa)[\bar f] + (\eta_a + \bar\eta_a)v^a,
	\end{align}
	where $\eta_a = - q_{ab} \nabla_{\Lb} L^b$ and $\bar\kappa = L_a \nabla_{\Lb} \Lb^a$. Since $h$ contains the transverse derivative $\Lb[\bar f]$, it ``talks to'' the extension of $\xi$ outside of $B$.
	\item $\LL_\xi q_{AB}$: We have $\delta q_{AB} = q_A{}^a q_B{}^b \delta g_{ab}$, hence, also using that $\theta_{AB} = \frac12 q_A{}^a q_B{}^b \L_L g_{ab}$,
	\begin{align}
		\LL_\xi q_{AB} = 2 f \theta_{AB} + \L_v q_{AB}.
	\end{align}
	Similarly,
	\begin{align}
		\LL_\xi q^{AB} = - 2 f \theta^{AB} + \L_v q^{AB}.
	\end{align}
	\item $\LL_\xi \varphi$, $\eb$ and $\es$: We have $\varphi = \frac{1}{D-2} \ln \sqrt q$, hence $\delta \varphi = \frac12 \frac{1}{D-2} q^{AB} \delta q_{AB}$. Using the previous we get
	\begin{align}
		\LL_\xi \varphi ={}& \frac{1}{D-2} q^{ab} \nabla_a (f L_b + \bar f \Lb_b + v_b)\cr
		={}& \frac{1}{D-2} (f \theta + \rd_A v^A).
	\end{align}
	Using $\delta \eb = (D-2) \delta \varphi \eb$, one gets
	\begin{align}
		\LL_\xi \eb = (f \theta + \rd_A v^A).
	\end{align}
	Using also $\L_\xi \eb = \rd \iota_\xi \eb = (f \theta + L[f] + \rd_A v^A) \eb$, which may be checked in a coordinate calculation, one gets
	\begin{align}
		\Delta_\xi \eb ={}& - L[f] \eb.
	\end{align}
	Using that $\Delta$ satisfies the Leibniz rule, we get
	\begin{align}
		\Delta_\xi (L^a \eb) ={}& 0,
	\end{align}
	thus as we claimed $L^a \eb$ is a density on $B$ valued into vectors of $B$ which is covariant under diffeomorphisms of $B$. Putting together the previous results, we have
	\begin{align}
		\LL_\xi (L^a \eb) = \L_\xi (L^a \eb) = [v, L]^a + (f \theta + \rd_A v^A) L^a \eb.
	\end{align}
	Since $\es = L^a \iota_a \eb$, we also get
	\begin{align}
		\Delta_\xi \es ={}& 0\cr
		\LL_\xi \es ={}& \L_\xi \es = \iota_{[v, L]} \eb + (f \theta + \rd_A v^A) \es.
	\end{align}
	Using $\L_\xi \es = \iota_\xi \rd \es + \rd \iota_\xi \es$, we have
	\begin{align}
		\iota_\xi \rd \es ={}& f \theta \es + \theta \iota_v \eb\cr
		\rd \iota_\xi \es ={}& \rd_A v^A \es + \iota_{[v, L]} \eb - \theta \iota_v \eb.
	\end{align}
	
	\item The derivative $\LL_\xi \gamma_{AB}$ may be derived using $\gamma_{AB} = e^{- 2\varphi} q_{AB}$ and the chain and Leibniz rules and reads
	\begin{align}
		\LL_\xi \gamma_{AB} = 2 f \sigma_{AB} - 2 e^{-2\varphi} \rd_{<A} v_{B>}.
	\end{align}
\end{itemize}
\textbf{Extrinsic Geometry:} The Weingarten map is the tensor $\nabla_a L^b$, which is a tensor on $B$ (i.e., the index $a$ is pulled back onto $B$ and the index $b$ is tangential to $B$).
Its transformation is easily worked out as
\begin{align}\label{eq:transfo_weingarten}
	\LL_\xi (\nabla_a L^b) \overset{B}{=}{}& \L_\xi (\nabla_a L^b) + \Delta_\xi (\nabla_a L^b) = \L_\xi (\nabla_a L^b) + \nabla_a (\Delta_\xi L^b)\cr
	={}& \L_\xi (\nabla_a L^b) + \nabla_a (L(f) L^b).
\end{align}
We used that the anomaly $\Delta_\xi$ commutes with the covariant derivative $\nabla$ as argued in section \ref{ssec:diffeos}, and plugged in the anomaly $\Delta_\xi L^a = L[f] L^a$.
The Weingarten map is thus non-covariant, but its transformation is independent of the extension of $\xi$. We will get the transformations of extrinsic geometry by taking components of the transformation of the Weingarten map. Recall that as a tensor on $B$,
\begin{align}
	\nabla_a L^b ={}& (\omega_a + \Lb_a \kappa) L^b + \theta_a{}^b.
\end{align}

\begin{itemize}
	\item $\kappa$ is the $\phi^0$-component of $L^a \nabla_a L^b$, thus
	\begin{align}
		\LL_\xi \kappa ={}& \Lb_b \LL_\xi (L^a \nabla_a L^b)\cr
		={}& \Lb_b [\L_\xi (L^a \nabla_a L^b) + (\Delta_\xi L^a) \nabla_a L^b + L^a \Delta_\xi (\nabla_a L^b)]\cr
		={}& v[ \kappa] + L[ (L + \kappa)[f]].
	\end{align}
	\item $\theta$: We have that $(\kappa + \theta)$ is the trace (on $B$) of the Weingarten map. Taking the trace of \Ref{eq:transfo_weingarten} gives:
	\begin{align}
		\LL_\xi (\kappa + \theta) ={}& \xi[\kappa + \theta] + L[L[f]] + L[f](\kappa + \theta)\cr
		={}&  v[\kappa + \theta] + L[ (L+\kappa+\theta)[f]].
	\end{align}
	Then using the result for $\kappa$, $\theta$ transforms as
	\begin{align}
		\LL_\xi \theta = v [\theta] + L[f \theta].
	\end{align}
	We get that $\theta \eb$ is covariant, which can also be seen because $\es$ is covariant and $\theta \eb = \rd \es$.
	\item $\mu$: combining the previous two results, we get
	\begin{align}
		\LL_\xi \mu = v[\mu] + L[ [L + \mu] f].
	\end{align}
	\item We have $\theta_{ab} = g_{bc} \nabla_a L^c$ (remember that everything is pulled back onto $B$), and using that $g_{ab}$ has vanishing anomaly and that $g(L)$ vanishes when pulled back,
	\begin{align}
		\LL_\xi \theta_{ab} = \L_\xi \theta_{ab} + L[f] \theta_{ab} = (f \L_L + L[f]) \theta_{ab} + \L_v \theta_{ab}.
	\end{align}
	Thus $\theta_{ab} \eb$ as a tensor on $B$ is covariant. The transformation of the upstairs extrinsic curvature and shear are obtained by combining with the transformation of $q^{AB}$ and $\varphi$. The upstairs extrinsic curvature transforms as
	\begin{align}
		\LL_\xi \theta^{AB} ={}& (f \p_0 + f \L_U + L[f]) \theta^{AB} + \L_v \theta^{AB}.
	\end{align}
	\item $\omega_A$: Note that $q_A{}^b = \p x^b/\p \sigma^A$ is independent of the metric, such that $\LL_\xi q_A{}^b = 0$. Also, $\Lb_a = (\rd \phi^0)_a$ as a tensor on $B$ is independent of the metric, so $\LL_\xi \Lb_a = 0$. With that,
	\begin{align}
		\LL_\xi \omega_A ={}& \LL_\xi (q_A{}^a \Lb_b \nabla_a L^b) = q_A{}^a \Lb_b \LL_\xi \nabla_a L^b)\cr
		={}& q_A{}^a \Lb_b (\L_\xi \nabla_a L^b + \nabla_a (L[f] L^b)) \cr
		={}& q_A{}^a \Lb_b \L_\xi (\omega_a L^b + \kappa \Lb_a L^b + \theta_a{}^b) + \p_A L[f] + L[f] \omega_A\cr
		={}& \L_v \omega_A + f q_A{}^a \L_L \omega_a + \p_A (\p_L f) + \kappa \p_A f - \theta_A{}^B \p_B f.
	\end{align}
	The same transformation has been given in \cite{Price:1986yy}.
\end{itemize}

\section{Relation of Hamiltonian and Noether Charge for Non-covariant Symplectic Potentials}\label{ap:ham}
The field space contraction of (the field space vector field induced by) the vector field $\xi$ with the symplectic form is calculated. We have
\begin{align}
	- I_\xi \Omega ={}& - \int_B I_\xi \delta \Theta\\
	={}& \int_B \delta (I_\xi \Theta) - \LL_\xi \Theta,
\end{align}
where we used the definition of the symplectic form on $B$, $\Omega = \int_B \delta \Theta$, and the Cartan formula $\LL_\xi = I_\xi \delta + \delta I_\xi$ for the field space Lie derivative. Next, use identity \Ref{eq:NT} for $I_\xi \Theta$, and the definition of the anomaly \Ref{eq:def_anom_on_forms}.
\begin{align}
	- I_\xi \Omega ={}& \int_B \delta (C_\xi + \iota_\xi L + \rd J_\xi) - \L_\xi \Theta - \Delta_\xi \Theta - I_{\delta \xi} \Theta.
\end{align}
Now use $\delta \iota_\xi L = \iota_\xi (-E + \rd \Theta) + \iota_{\delta \xi} L$, and $\L_\xi \Theta = \iota_\xi \rd \Theta + \rd \iota_\xi \Theta$, and $I_{\delta \xi} \Theta = C_{\delta \xi} + \iota_{\delta \xi} L + \rd J_{\delta \xi}$. Setting all constraint  terms to zero, we get
\begin{align}
	-I_\xi \Omega \hat ={}& \int_B - \Delta_\xi \Theta + \int_{\p B} \delta (J_\xi) - J_{\delta \xi} - \iota_\xi \Theta.
\end{align}
Since the standard symplectic potential current $\hat \Theta$ is covariant, in our case the anomaly of $\Theta$ comes from the boundary modification: $\Delta_\xi \Theta = \rd (\Delta_\xi \alpha_S)$, so we get the equation \Ref{eq:ham_for_anomalous} used in the main text:
\begin{align}
	-I_\xi \Omega \hat ={}& \int_{\p B} \big( \delta (J_{\xi}) - J_{\delta \xi} - \iota_\xi \Theta - \Delta_\xi \alpha_S \big).
\end{align}

We also need the anomaly of the modification $\alpha_S = \frac12 L_a \delta g^{ab} \iota_b \eb + \delta L^a \iota_a \eb - \frac{1}{D-2}\delta \es$. The anomalies of $\delta \es$, of $\delta g^{ab}$ and of $L_a \eb$ vanish for $\xi \parallel B$, so we are left with
\begin{align}
	\Delta_\xi \alpha_S ={}& (\LL_\xi - \L_\xi - I_{\delta \xi}) (\delta L^a \iota_a \eb)\\
	={}& (\Delta_\xi \delta L^a) \iota_a \eb + \delta L^a \iota_a (\Delta_\xi \eb),
\end{align}
where for the second line we have used that $\Delta_\xi$ satisfies the Leibniz rule. Now use that $\delta \Delta_\xi = \Delta_\xi \delta + \Delta_{\delta \xi}$, which follows from its definition. Using the results $\Delta_\xi L^a = L[f] L^a$ and $\Delta_\xi \eb = -L[f] \eb$, as well as $\Delta_{\delta \xi} L^a = L[\delta f] L^a$, one obtains
\begin{align}
	\Delta_\xi \alpha_S ={}& (\delta L)[f] \es,
\end{align}
where as before $\xi = f L + v$ with $v \parallel S$.

\bibliographystyle{uiuchept}
\bibliography{references}
\end{document}